\documentclass[fleqn,usenatbib]{mnras}

\usepackage[T1]{fontenc}

\DeclareRobustCommand{\VAN}[3]{#2}
\let\VANthebibliography\thebibliography
\def\thebibliography{\DeclareRobustCommand{\VAN}[3]{##3}\VANthebibliography}

\usepackage{graphicx}	
\usepackage{amsmath}	
\usepackage{amssymb}	
\usepackage{hyperref}

\def \aj {AJ}
\def \mnras {MNRAS}
\def \apj {ApJ}
\def \apjs {ApJS}
\def \apjl {ApJL}
\def \aap {A\&A}
\def \nat {Nature}

\def \pasp {PASP}

\usepackage{newtxtext,newtxmath}

\title[The radio counterpart of LS V +44 17]{VLA monitoring of LS V +44 17 reveals scatter in the X-ray -- radio correlation of Be/X-ray binaries}

\author[J. van den Eijnden et al.]{J. van den Eijnden,$^{1}$
A. Rouco Escorial,$^{2}$
J. Alfonso-Garz\'on,$^{3}$
J. C. A. Miller-Jones,$^{4}$
P. Kretschmar,$^{2}$
\newauthor F. F\"urst,$^{5}$
N. Degenaar,$^{6}$
J. V. Hern\'andez Santisteban,$^{7}$
G. R. Sivakoff,$^{8}$
T. D. Russell,$^{9}$
R. Wijnands$^{6}$
\\
$^{1}$Department of Physics, University of Warwick, Coventry CV4 7AL, UK\\
$^{2}$ European Space Agency (ESA), European Space Astronomy Centre (ESAC), Camino Bajo del Castillo s/n, 28692, Villanueva de la Cañada, Madrid, Spain\\
$^{3}$ Centro de Astrobiolog\'{\i}a-Departamento de Astrof\'{\i}sica (CSIC-INTA), Camino Bajo del Castillo s/n,
28692 Villanueva de la Ca\~nada, Spain\\
$^{4}$ International Centre for Radio Astronomy Research, Curtin University, GPO Box U1987, Perth, WA 6845, Australia\\
$^{5}$ Quasar Science Resources SL for ESA, European Space Astronomy Centre (ESAC), Science Operations Departement, 28692, Villanueva de la Cañada, Madrid, Spain\\
$^{6}$ Anton Pannekoek Institute for Astronomy, University of Amsterdam, Science Park 904, Amsterdam, 1098 XH, The Netherlands\\
$^{7}$ SUPA Physics and Astronomy, University of St Andrews, Scotland KY16 9SS, UK\\
$^{8}$ Department of Physics, University of Alberta, CCIS 4-181, Edmonton AB T6G 2E1, Canada\\
$^{9}$ Istituto di Astrofisica Spaziale e Fisica Cosmica, INAF, Via U. La Malfa 153, Palermo, I-90146, Italy\\
}

\date{Accepted XXX. Received YYY; in original form ZZZ}

\pubyear{2023}

\begin{document}
\label{firstpage}
\pagerange{\pageref{firstpage}--\pageref{lastpage}}
\maketitle

\begin{abstract}
LS V +44 17 is a persistent Be/X-ray binary (BeXRB) that displayed a bright, double-peaked period of X-ray activity in late 2022/early 2023. We present a radio monitoring campaign of this outburst using the Very Large Array. Radio emission was detected, but only during the second, X-ray brightest, peak, where the radio emission followed the rise and decay of the X-ray outburst. LS V +44 17 is therefore the third neutron star BeXRB with a radio counterpart. Similar to the other two systems (Swift J0243.6+6124 and 1A 0535+262), its X-ray and radio luminosity are correlated: we measure a power law slope $\beta = 1.25^{+0.64}_{-0.30}$ and a  radio luminosity of $L_R = (1.6\pm0.2)\times10^{26}$ erg/s at a $0.5-10$ keV X-ray luminosity of $2\times10^{36}$ erg/s (i.e. $\sim 1\%$ $L_{\rm Edd}$). This correlation index is slightly steeper than measured for the other two sources, while its radio luminosity is higher. We discuss the origin of the radio emission, specifically in the context of jet launching. The enhanced radio brightness compared to the other two BeXRBs is the first evidence of scatter in the giant BeXRB outburst X-ray -- radio correlation, similar to the scatter observed in sub-classes of low-mass X-ray binaries. While a universal explanation for such scatter is not known, we explore several options: we conclude that the three sources do not follow proposed scalings between jet power and neutron star spin or magnetic field, and instead briefly explore the effects that ambient stellar wind density may have on BeXRB jet luminosity.
\end{abstract}


\section{Introduction}

X-ray binaries, wherein a compact object accretes mass from an orbiting companion star, are prolific radio sources due to the launch of jets from their accretion flow or spinning black hole. However, a substantial sub-class of X-ray binaries, the high mass X-ray binaries (HMXBs), have only emerged as radio emitters in the past decade. HMXBs host a massive ($>10$ $M_{\odot}$) early-type star as their donor, while the compact object is nearly always a slow pulsar: a neutron star with a strong magnetic field \citep[$B\geq10^{12}$ G;][]{staubert2019} and spin period typically longer than $\sim 1$ second \citep[see e.g.][for a recent catalogue]{fortin2023}. Only a small number of HMXBs, such as Cyg X-1, likely Cyg X-3 \citep{vankerkwijk1992}, and possibly MWC 656 \citep{casares2014,ribo2017}, are known to host black holes; Cyg X-1 and Cyg X-3 are bright radio emitters when actively accreting\footnote{Radio observations of MWC 656 have only occurred in its quiescent state \citep{ribo2017}. Also,  recent works have argued that MWC 656 does not host a black hole as the Be-star's companion, but instead a hot subdwarf \citep{rivinius2022} or a neutron star, white dwarf, or hot helium star \citep{janssens2023}.}. With increased sensitivity in the radio band, the more common neutron star HMXBs are now also often detected at these low frequencies \citep{vandeneijnden2021,vandeneijnden2022}. 

The radio properties of neutron star HMXBs depend on the sub-class of HMXB considered. The donor stellar type often drives this classification, firstly between OB supergiant and Be-star systems. The former, where the neutron star orbits closely in the strong stellar wind of the supergiant, accrete either transiently or persistently (around $L_X \sim 10^{35} - 10^{36}$ erg/s); a difference that remains poorly understood \citep[see e.g.][for a recent review]{Sidoli2017}. In this work, we will focus instead on the Be systems \citep[see][for an overview]{reig2011}. Be stars are late O-type/early B-type stars with luminosity class III-V, displaying emission lines and IR excesses that are attributed to a decretion (also known as circumstellar) disc around the star. HMXBs hosting a Be donor are predominantly transient, accreting either at periastron passage as the neutron star passes through the Be-star’s circumstellar disc or at other orbital phases when a giant outburst is triggered via a still poorly understood mechanism \citep{okazaki01,moritani13,martin14,monageng17,laplace17,martin2019}. 

A noteworthy complication to this picture, however, is that a small but significant subset of this Be/X-ray binary (BeXRB) class persistently accretes at low X-ray luminosities \citep[$L_X \sim 10^{34} - 10^{35}$ erg/s;][]{reig1999,tsygankov2017b} in between their brighter periods of activity, instead of returning to quiescence. As discussed below, the target of this paper is one of these systems. 

To date, two BeXRBs have been detected in the radio band: Swift J0243.6+6124 \citep[during two outbursts;][]{vandeneijnden2018,vandeneijnden2023} and 1A 0535+262 \citep{vandeneijnden2022}. When tracking their X-ray and radio luminosity quasi simultaneously, a correlation is observed, which appears to be consistent between both sources. Such X-ray -- radio correlations with different characteristics are also seen in black hole and neutron star low-mass X-ray binaries (LMXBs), attributed to the coupling between the X-ray emitting inflow and radio-emitting jet \citep{hannikainen98,corbel03,gallo03,migliari06,tudor17}. Importantly, the radio luminosities seen in giant BeXRB outbursts are orders of magnitude lower at the same X-ray luminosity than in LMXBs \citep{migliari06,gallo18,vandeneijnden2022}. In BeXRBs, the radio emission has similarly been interpreted as jet emission \citep[see e.g.][for detailed discussions on this origin and alternative interpretations, such as colliding disk and stellar winds]{vandeneijnden2021,chatzis2022}. While the radio coverage of the giant outbursts of Swift J0243.6+6124 and 1A 0535+262 does not overlap in X-ray luminosity, their X-ray -- radio correlations are consistent with lying on a single relation. This consistency, albeit in merely two sources, is in contrast with LMXBs: both black hole and neutron star LMXBs show significant scatter and, for black holes, evidence for two tracks in the X-ray -- radio luminosity plane \citep[e.g.][]{coriat2011,gallo2014,gallo18,motta2018}.

This BeXRB jet interpretation then introduces a number of further questions: what mechanism underlies the launch of these jets, especially given the strong magnetic field of the neutron star and its disruptive effects on the inner accretion disc \citep{massi2008, parfrey2016, das2022}? What powers BeXRB jets and explains their radio faintness? Do all giant BeXRBs outbursts follow a single inflow--outflow coupling or does this population contain scatter similar to LMXBs, caused by additional factors -- neutron star magnetic field and spin, binary orbital properties, ambient stellar wind density --- that play a role in regulating the jet’s luminosity? And how does giant outburst radio emission compare to other types of BeXRB activity, such as periastron outbursts and X-ray re-brightenings?

In this paper, we present a radio monitoring campaign of a third radio-detected neutron star BeXRB. We study LS V +44 17 (also known as RX J0440.9+4431), a BeXRB located at a 
distance of $2.44 \pm 0.10$ kpc \citep[derived from Gaia data by][]{fortin2023}, with a $\sim 205$-second neutron star spin \citep{lapalombara2012} and a B.2Ve donor star \citep{reig2005}, that resides in a $150$-day orbit \citep{ferrigno2013}. LS V +44 17 is an example of a persistent BeXRB, but also showed periods of enhanced activity in 1984, 1997, and 2010--2011 \citep{reig1999,morii2010} --- suggestive of a $\sim 13$-year super-orbital cycle in this activity that is also consistent with its 2023 outburst. Detailed analysis of the three periods of X-ray enhancement in 2010--2011 suggested they were caused by periastron passages \citep{morii2010,finger2010,tsygankov2011}. The persistent activity level, typically a couple times $10^{34}$ erg/s, was originally attributed to ongoing low-level accretion from a weak Be-star wind \citep{reig1999,ferrigno2013}, but was more recently attributed to accretion from a cold (neutral) accretion disc: a configuration that may manifest around neutron stars spinning too slowly to enter the propeller regime \citep{illarionov1975} before the disc cools into its neutral state \citep{tsygankov2017,rouco2020,sagalnik2023b}.

\begin{figure*}
\begin{center}
\includegraphics[width=\textwidth]{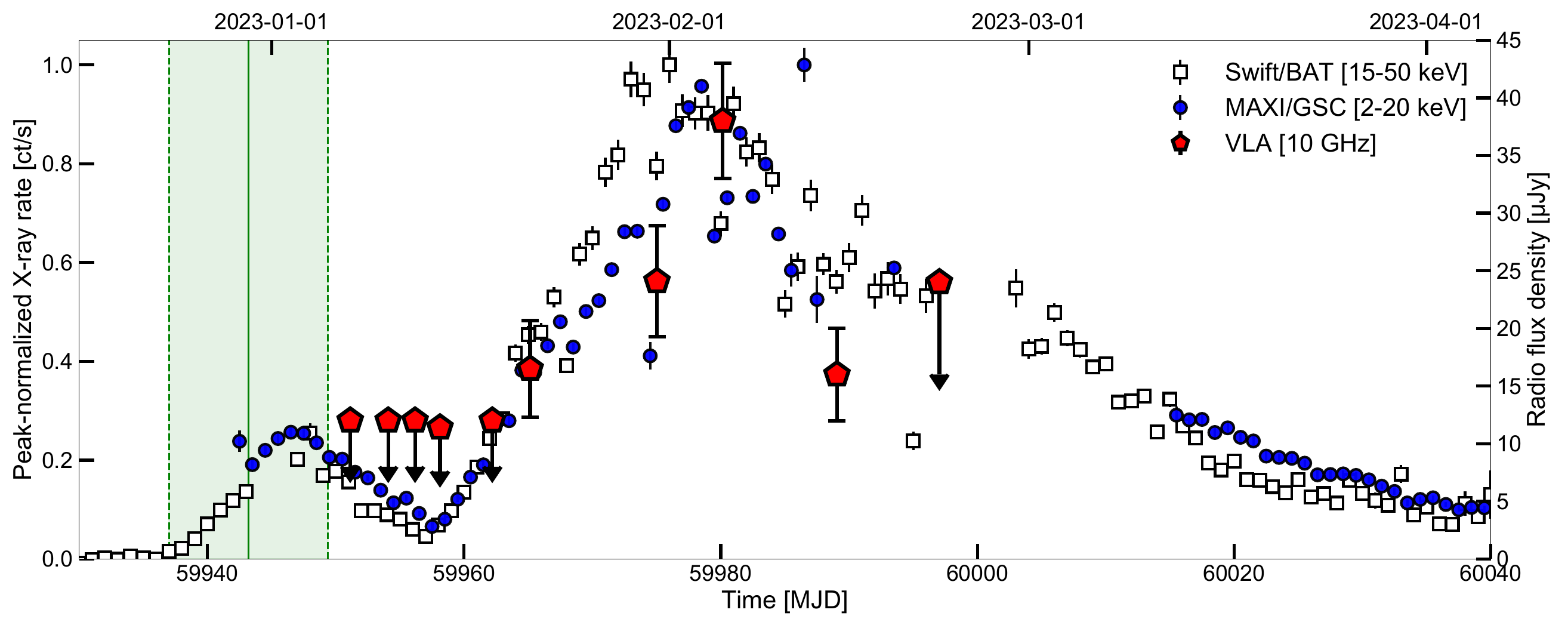}
\caption{X-ray and radio light curves of the 2022/2023 outburst of LS V +44 17. The daily X-ray monitoring light curves from \textit{MAXI}/GSC and \textit{Swift/BAT}are shown in the blue circles and white squares, respectively, normalised to their maximum count rate during the outburst (left y-axis). The red pentagons show the 10-GHz (e.g. X-band) VLA flux densities (right y-axis). Two peaks are visible during the X-ray outburst, while the radio counterpart only becomes detectable during the second, X-ray-brighter peak. Despite the lower VLA cadence, the radio counterpart can be seen to brighten with X-ray count rates during the second peak, falling again after the X-rays have reached their maximum. The green line and shaded area indicate periastron passage and its uncertainty (from extrapolating the period uncertainty) based on \citet{ferrigno2013}.}
\label{fig:lcs}
\end{center}
\end{figure*}

In late 2022/early 2023, LS V +44 17 showed its X-ray brightest ever recorded activity period \citep{nakajima2022, mandal2023} with a complex, double-peaked outburst profile \citep[][see also Figure \ref{fig:lcs}]{pal2023,palmer2023,coley2023,salganik2023,gaishin2023}. Such complex outburst profiles have been reported for other Be/X-ray binaries previously, including in 1A 0535+262 \citep{caballero2013, moritani13}, GX 304-1 \citep{postnov2015}, and GRO J1008-57 \citep{kuhnel2017}. Such rapid consecutive periods of activity, often during a single orbital period, are often associated with a warped and precessing circumstellar disc around the Be-star \citep{moritani13,martin14,okazaki2013,franchini2019}. The 2022/2023 outburst of LS V +44 17 was monitored across a wide range of wavelengths, by observatories including VERITAS at $\gamma$-ray energies \citep[][not detected]{veritas2023} and the Imaging X-ray Polarimetry Explorer \citep[IXPE;][]{doroshenko2023}, which can provide unique measurements of the neutron star’s magnetic field configuration (e.g. the relative angles between the neutron star spin, magnetic field, and disc spin axes) via X-ray polarimetry \citep[see also e.g.][]{mushtukov2023}. We monitored and detected this giant outburst with the Karl G. Jansky Very Large Array (VLA) in the radio band. In this work, we report on the setup, outcome, and implications of this monitoring campaign. 

\section{Observations and data reduction}

We performed ten observations with the VLA between 2023 January 7 and February 22 (program VLA/22B-051), with a separation between observations ranging between $2$ and $10$ days. Seven of these observations lasted $2$ hours including overheads (1 hour and 16 minutes on target), in comparison to $1.5$ hours (50 minutes on target) for the remaining three. Regardless of total observing time, the integration time for each observation was divided in half for the C and X bands, centred at $6$ and $10$ GHz, respectively, in $3$-bit mode to provide full spectral coverage between $4$ and $12$ GHz. As primary and secondary calibrators, we observed 3C 48 at the start of the observations and J0440+427 interleaved between target scans, respectively. Target scans of LS V +44 17 were offset by $10$" North to prevent phase centre image artifacts at the target position. During the first three observations, the array was in the C$\rightarrow$B configuration, switching to the B configuration afterwards. We performed all further data reduction using the \textsc{Common Astronomy Software Applications} package \citep[\textsc{CASA};][]{casa2022}, flagging the data using a combination of automatic routines and manual inspection, before performing calibration and imaging. To balance resolution and the suppression of sidelobes, we applied Briggs weighting with a robust parameter of $0$ and $1$ for C and X band, respectively. Whenever a radio source was detected at $>3\sigma$ significance consistent with the position of LS V +44 17, we applied the \textsc{CASA} task \textsc{imfit} to fit a 2D Gaussian profile matching the synthesized beam in full-width half maximum and orientation and measure the flux density. Finally, we measured the RMS sensitivity of each observation as the RMS over a nearby, source-free region, or over the target region in case of a non-detection. Information about each observation and the resulting flux density measurements are listed in Table \ref{tab:data}. We note that an additional VLA Director's Discretionary Time observation was performed on 17 Feb 2023 \citep[VLA/23A-384;][]{kumari2023}, not detecting any emission, which we do not analyse further in this work\footnote{The short total exposures per band (L, C, X) yield sensitivities that are not constraining compared to the flux densities reported in Section 3.}. 

The outburst of LS V +44 17 was monitored extensively in X-rays by a suite of observatories, including the \textit{Neutron Star Interior Composition Explorer} \citep[\textit{NICER};][]{gendreau2016} through publicly available observations. \textit{NICER} observed LS V +44 17 almost daily during the period of radio monitoring, providing X-ray observations quasi-simultaneous with each VLA observation (program ID 5203610). This X-ray monitoring campaign is therefore perfectly suited for a comparative X-ray and radio study. We acquired the \textit{NICER} observations closest to the ten VLA observations (nine taken on the same MJD, the final \textit{NICER} data collected a day later; see Table \ref{tab:data}) from the \textsc{heasarc} and used the standard tools in \textsc{nicerdas} v10 from \textsc{HEASoft} v6.31.1 with calibration files from CALDB xti20221001 to create spectra and responses. Specifically, we used the \textsc{nicerl2} and \textsc{nicerl3-spec} tools with the 3c50 background model \citep{remillard2019}. Pointed observations of LS V +44 17 were also performed by the Neil Gehrels \textit{Swift} Observatory \citep[\textit{Swift};][]{gehrels2004} and the \textit{Nuclear Spectroscopic Telescope Array} \citep[\textit{NuSTAR};][]{harrison2012}. However, the \textit{Swift} monitoring, with a typical cadence of one observation per three days, is sparser than the \textit{NICER} monitoring and therefore leaves larger gaps from the VLA radio observations. \textit{NuSTAR} observed three times \citep[see e.g.][]{sagalnik2023b}, but none of these observations were taken within a day from a radio observation. We also note that the \textit{Imaging X-ray Polarimetry Explorer} \citep[\textit{IXPE};][]{IXPEref} observed the target twice; while we do not analyse these observations here, we will compare their results \citep[e.g.][]{doroshenko2023} with our radio monitoring.

Finally, the overall X-ray outburst profile was monitored at high cadence by the \textit{Swift} / Burst Alert Telescope \citep[BAT;][]{batreference,krimm2013} and the \textit{Monitor of All-sky X-ray Image} (\textit{MAXI}) / Gas Slit Camera \citep[GSC;][]{maxiref}. We extracted the daily source light curves from their respective light curve repositories for known X-ray sources\footnote{\url{http://maxi.riken.jp/star\_data/J0440+445/J0440+445.html} and \url{https://swift.gsfc.nasa.gov/results/transients/weak/LSVp4417/}.}.

\section{Results}

\subsection{The radio counterpart of LS V +44 17}
\label{sec:radiodet}

In Figure \ref{fig:lcs}, we show the X-ray and radio light curves for the late 2022 / early 2023 activity of LS V +44 17. The X-ray outburst started around MJD 59940, initially rising for ${\sim}10$ days, peaking at a time consistent with periastron passage \citep{ferrigno2013}, followed by a decay on a similar time scale. Before fully decaying into quiescence, however, a second rise set in, eventually reaching a ${\sim}4$ times higher flux than the initial outburst peak approximately 20 days later \citep{ pal2023,palmer2023,coley2023,salganik2023,gaishin2023}. After this second peak, the source steadily decayed back towards quiescent levels over a time scale of ${\sim}2$ months. During the second peak of the outburst, both the \textit{Swift}/BAT and \textit{MAXI}/GSC light curves show day-to-day variability superimposed upon the mean outburst profile. These variations are expected due to the long, $\sim 205$ second pulse period of LS V +44 17, in combination with the large pulse fraction of BeXRBs in outburst and the short individual exposures per scan of the two X-ray monitors\footnote{\textit{Swift}/BAT individual scans are limited by \textit{Swift}'s observing schedule, which observes a particular field for continuous exposures up to $\sim 1200$ seconds at a time \citep{krimm2013}. \textit{MAXI}/GSC visits each field for $40$-$150$ seconds at a time. A similar but stronger beating effect is observed and modelled in detail in \citet{pike2023} for the 2022 outburst of the BeXRB MAXI J0655-013.}. As expected, for LS V +44 17, these beating effects between the neutron star pulse and observing cadence become more apparent in their higher time resolution light curves (which are not shown in Figure \ref{fig:lcs}, but are available in the observatory's online light curve repositories).

\begin{table*}
\caption{Details of the radio and X-ray observations analysed in this work. For each observation, we list the observation number, the VLA start date in MJD, VLA observation length $T_{\rm obs}$ in hours, radio observing frequency $\nu$ in GHz, measured radio flux density $S_\nu$ or $3$-$\sigma$ upper limit (indicated with "<") in $\mu$Jy, and the VLA configuration. All radio observations were performed under program VLA/22B-051. We also list the \textit{NICER} ObsID for the accompanying X-ray observation, its MJD, and unabsorbed 0.5--10 keV X-ray flux. Fitting all spectra jointly, the absorption column was found to be $N_H = (5.00\pm0.18)\times10^{21}$ cm$^{-2}$. $^{a}$We note that for each \textit{NICER} ObsID, all individual exposures on the associated MJD are combined into a single spectrum. }
\label{tab:data}
\begin{tabular}{lllllllll}
\hline
Obs. No. & VLA MJD & $T_{\rm obs}$ [hr] & Config. & $\nu$ [GHz] & $S_\nu$ [$\mu$Jy] & \textit{NICER} ObsID & \textit{NICER} MJD$^{a}$ & 0.5--10 keV flux [erg/s/cm$^{2}$] \\
\hline
1 & 59951.13 & 2.0 & C$\rightarrow$B & 6.0 & $< 17.0$ & 5203610110 & 59951 & $(3.319\pm 0.008)\times10^{-9}$ \\
& & & & 10.0 & $< 12.0$  \\ \hline
2 & 59954.08 & 2.0 & C$\rightarrow$B & 6.0 & $< 18.0$  & 5203610113 & 59954 & $(2.372\pm 0.008)\times10^{-9}$ \\
& & & & 10.0 & $< 12.0$  \\ \hline
3 & 59956.18 & 2.0 & C$\rightarrow$B & 6.0 & $< 24$  & 5203610115 & 59956 & $(1.648\pm 0.007)\times10^{-9}$ \\
& & & & 10.0 & $< 12.0$  \\ \hline
4 & 59958.12 & 2.0 & B & 6.0 & $< 27.0$  & 5203610117 & 59958 & $(1.646\pm 0.007)\times10^{-9}$ \\
& & & & 10.0 & $< 11.4$  \\ \hline
5 & 59962.20 & 2.0 & B & 6.0 & $< 22.5$  & 5203610121 &59962 & $(4.807\pm 0.017)\times10^{-9}$ \\
& & & & 10.0 & $< 12.0$  \\ \hline
6 & 59965.14 & 2.0 & B & 6.0 & $< 24$  & 5203610124 & 59965 & $(7.148\pm 0.020)\times10^{-9}$ \\
& & & & 10.0 & $16.5 \pm 4.2$  \\ \hline
7 & 59975.03 & 1.5 & B & 6.0 & $< 22$  & 5203610134 & 59975 & $(1.434\pm 0.023)\times10^{-8}$ \\
& & & & 10.0 & $24.1 \pm 4.8$  \\ \hline
8 & 59980.14 & 1.5 & B & 6.0 & $26.3 \pm 8.0$  & 5203610139 & 59980 & $(1.356\pm 0.021)\times10^{-8}$ \\
& & & & 10.0 & $38.0 \pm 5.0$  \\ \hline
9 & 59989.05 & 2.0 & B & 6.0 & $22.7 \pm 6.5$  & 5203610148 & 59989 & $(1.128\pm 0.004)\times10^{-8}$ \\
& & & & 10.0 & $16.0 \pm 4.0$ \\ \hline
10 & 59997.03 & 1.5 & B & 6.0 & $< 24$  & 5203610155 & 59998 & $(9.155\pm 0.050)\times10^{-9}$ \\
& & & & 10.0 & $< 24$  \\ \hline
\end{tabular}\\
\end{table*}


The first four radio observations took place during the decay of the first outburst peak, yielding four non-detections at both C (6 GHz) and X (10 GHz) band. The same holds for the fifth observation, during the second peak's rise, after which LS V +44 17 is first detected at an X band flux density of $S_\nu = 16.5 \pm 4.2$ $\mu$Jy. It is noteworthy that this first radio detection occurs when the X-ray flux first rises above the peak of the first outburst, consistent with a link between the transient X-ray and radio emission given the sensitivity limits underlying the earlier radio non-detections. Following the X-ray lightcurve, which peaks between MJD 59975-59980, the radio light curve rises after MJD 59965 before it peaks during Observation 8 (MJD 59980) at $S_\nu = 38.0 \pm 5.0$ $\mu$Jy in X band, before finally decaying until the source is not detected in the final radio observation. We note that the final radio observation suffered significantly lower X-band sensitivity due to enhanced RFI (8 $\mu$Jy vs. $\sim$4 $\mu$Jy), which would not have yielded a source detection in either the sixth or ninth observation. All radio flux densities are listed in Table \ref{tab:data}.

\begin{figure*}
\begin{center}
\includegraphics[width=\textwidth]{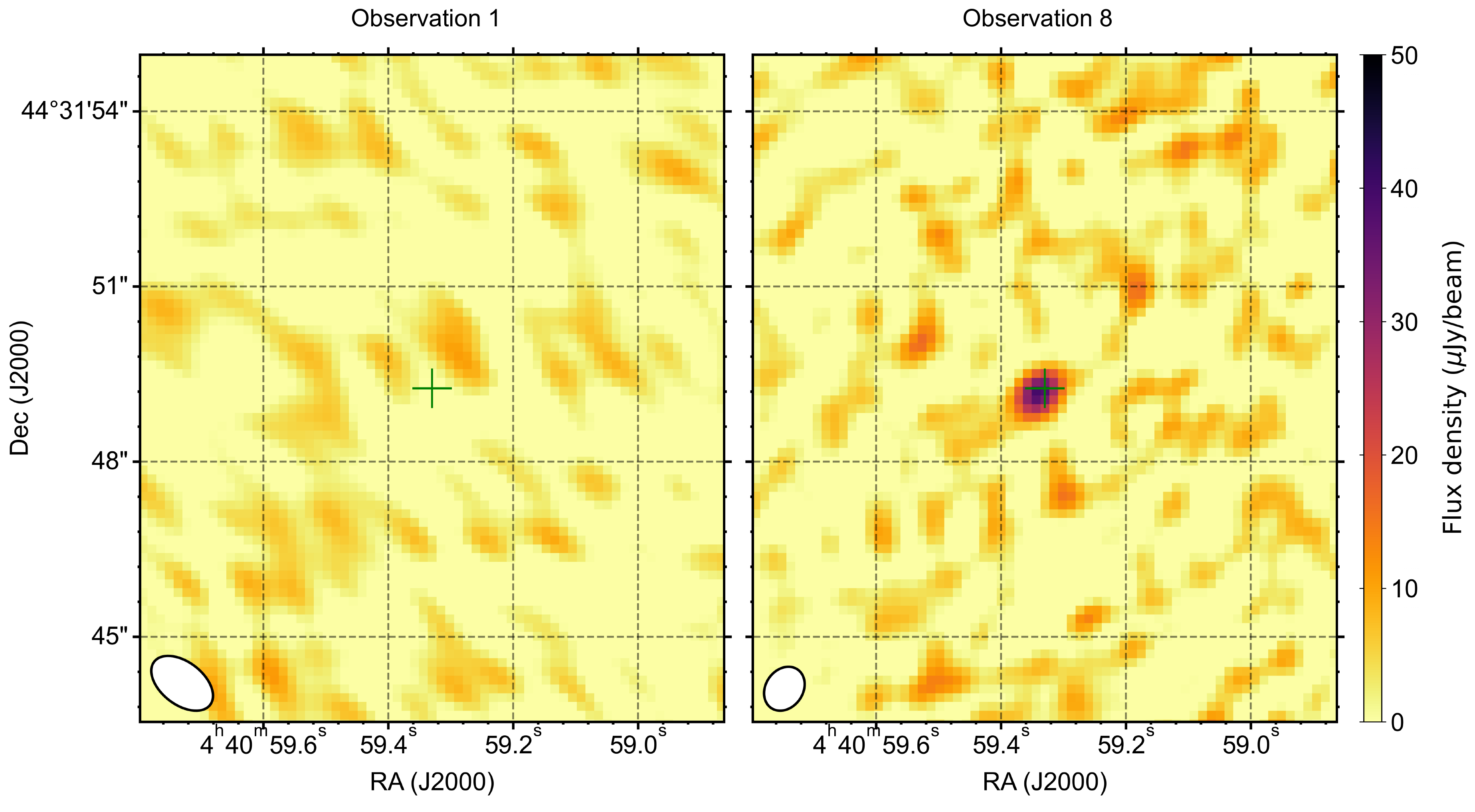}
\caption{The inner $10\times10$ arcsecond region centered at the Gaia position of LS V +44 17, as seen by the VLA at 10 GHz. The left panel shows the first VLA observation, fully consistent with noise, while the right panel shows how a radio counterpart has appeared in the outburst's second X-ray peak (seen in Observations 6 to 9, here shown during Observation 8). The color scaling is the same in both panels. The synthesized beams for both observations are shown in bottom left, changing size due to a configuration change after the third observation. We note that the phase-center of the observations was pointed 10 arcseconds North of LS V +44 17, i.e. located 5 arcseconds above both panels.}
\label{fig:radioimages}
\end{center}
\end{figure*}

Figure \ref{fig:radioimages} shows a $\sim 10\times10$ arcsecond zoom of the 10-GHz band field of view around the target position, comparing the first (left) and eighth (right; the radio brightest) VLA observation. In both images, the green cross shows the Gaia position of LS V +44 17. A clear radio counterpart is present in Observation 8, while in Observation 1, the entire zoom image is consistent with background noise. All four radio detections were obtained in the same array configuration and at a S/N ratio less than 10, 
implying that we are limited by statistical rather than systematic uncertainties, and therefore the accuracy the accuracy of each observation's radio position measurement is given by the beam size divided by the S/N. Therefore, we measure the most accurate radio position, consistent with the most accurate known optical position from Gaia, in Observation 8:

\smallskip

\smallskip

\noindent RA = $04$h $40$m $59.34$s $\pm$ $0.01$s

\noindent Dec = $+44^{\circ}$ $31$' $49.16$" $\pm$ $0.10$"

\smallskip

\smallskip

\noindent At C band, the sensitivity of our observations is typically a factor $\sim$two lower. Combined with the limited S/N ratio of the X band detections, C band detections are therefore only expected for observations with steep spectral shapes or bright X band flux densities. Indeed, we obtain C band detections only in observations 8 and 9. In both cases, as shown in Table \ref{tab:data}, the counterparts are seen at a significance slightly above 3 $\sigma$. In both cases, the position of the source is consistent with the X band detection within 1 $\sigma$ uncertainties. Comparing both bands, we measure spectral indices of $\alpha = 0.7 \pm 1.5$ and $\alpha = -0.7 \pm 1.7$, where $S_\nu \propto \nu^{\alpha}$. Due to the relatively low signal to noise ratio, both measurements are consistent but poorly constrained and do not distinguish between steep non-thermal (where $\alpha$ is negative) or inverted ($\alpha > 0$) broad-band spectral shapes.

\subsection{X-ray flux measurements}

We performed X-ray spectral analysis to measure the X-ray luminosity as close to the time of the radio observations as possible. In all but one observation (VLA Observation 10), \textit{NICER} X-ray observations were available on the same day as the VLA observation. The bandpass of \textit{NICER} fully covers the 0.5--10 keV energy range used typically in X-ray binaries as the tracer for the accretion luminosity\footnote{After our spectral analysis, we confirmed that the limited bandpass of \textit{NICER} did not introduce systematic biases in the flux determination: the 0.5--10 keV fluxes extrapolated from the models fitted to the \textit{NuSTAR} data by \citep{sagalnik2023b}, agreed at the $\sim$per cent level with the fluxes we measured in that band with \textit{NICER} on the same days, using our approach.}. We performed all fits described below using \textsc{xspec} v.12.13.0c \citep{xspecref}, assuming interstellar abundances from \citet{wilms2000} and cross-sections from \citet{vern1996}. For each attempted model, all spectra were fit jointly in \textsc{xspec}. In those fits, all model parameters were left to vary between the spectra, with on exception: in considered models, we included interstellar absorption via the \textsc{tbabs} model, and kept the absorbing column density $N_H$ tied between all spectra. All considered models are discussed in the next paragraph.

Trial fits to the \textit{NICER} spectra with single-component models, i.e. absorbed thermal or non-thermal spectra, failed to provide an adequate description of the data. In addition to showing the need for a two-component continuum, such trials showed that all attempted models require the inclusion of a iron line feature around $6.4$ keV, which we modeled as a Gaussian line. The broadband X-ray spectra of LS V +44 17 have previously been fitted with different composite models: an absorbed blackbody plus cutoff power law model, used by \citet{tsygankov2012}, and an absorbed double \textsc{comptt} model, used by \citet{sagalnik2023b}. The latter model, where the highest-energy \textsc{comptt} component dominates above the \textit{NICER} bandpass, was found to perform better across the $1$-$79$ keV band \citep{sagalnik2023b}. However, below $10$ keV, the former model provides an adequate description of the spectra. We therefore fit the ten \textit{NICER} spectra jointly with a \textsc{tbabs*(bbody + cutoffpl + gauss)} model with tied absorption, finding a reduced $\chi_\nu^2 = 1164.0/1353$. All fit parameters are listed in Appendix \ref{app}; we find an absorption column of $N_H = (5.00\pm0.18)\times10^{21}$ cm$^{-2}$, close to the Galactic value in the target direction \citep[$\sim 6\times10^{21}$ cm$^{-2}$;][]{2016A&A...594A.116H}. We then measured the unabsorbed $0.5$-$10$ keV flux using the convolution model \textsc{cflux}. All fluxes are listed in Table \ref{tab:data}. Finally, for comparison, we also fitted the spectra replacing the cutoff power law with a normal power law, finding a worse fit ($\Delta \chi^2 \approx +98.0$ for 10 fewer parameters\footnote{Note that we cannot perform an f-test to calculate the significance of including the cutoff as the more complex model is not created by adding a component to the more simple model.}). 

\subsection{The X-ray -- radio luminosity plane}
\label{sec:lxlrcor}

\begin{figure*}
\begin{center}
\includegraphics[width=\textwidth]{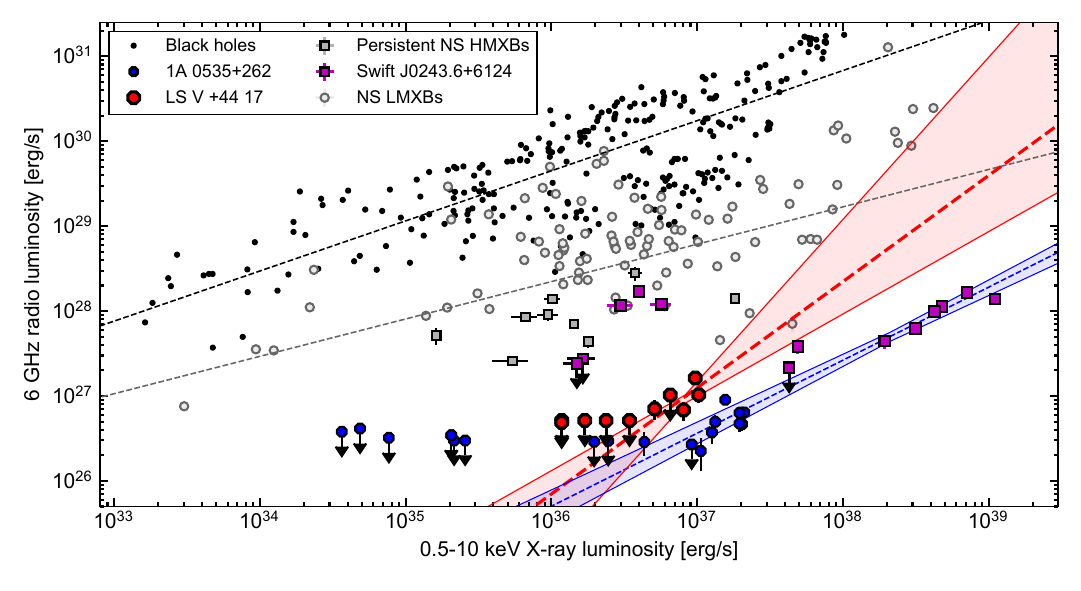}
\caption{The X-ray -- radio luminosity plane for low-mass and high-mass X-ray binaries, highlighting that the X-ray -- radio coupling of LS V +44 17 is radio brighter than that of the other BeXRBs but remains faint compared to LMXBs. Three neutron star BeXRBs are shown: LS V +44 17 as the red octagons, complemented by Swift J0243.6+6124 (purple squares; note that all datapoints below $L_X = 10^{37}$ erg were obtained during an X-ray re-brightening instead of a giant outburst), and 1A 0535+262 as the blue octagons. For clarity, we do not show the radio-non-detected BeXRB GRO J1008-57. The red line and pink shaded area show the best X-ray -- radio correlation fit, with $1$-$\sigma$ uncertainties, to the data of LS V +44 17. The blue dashed line and light blue region indicates the best fit correlation to the combined datasets of the other BeXRBs \citep[][including the radio non-detected GRO J1008-57 but excluding the re-brightening of Swift J0243.6+6124; see Section \ref{sec:41}]{vandeneijnden2022}. For comparison, black hole and neutron star LMXBs are shown as the black points and grey circles, respectively (based on Gallo et al. 2018). SgXBs are shown as grey squares, based on \citet{vandeneijnden2021}, showing only radio-detected sources for clarity.}
\label{fig:lxlr}
\end{center}
\end{figure*}

We assumed a distance of $2.44 \pm 0.10$ kpc, derived by \citet{fortin2023} from Gaia EDR3 \citep{Bailer-Jones2021}, to convert the X-ray fluxes and radio flux densities to luminosities. We convert the X-band radio measurements to $6$ GHz luminosities assuming a flat spectrum, given the lack of constraining spectral index information (Section \ref{sec:radiodet}). We plot the resulting luminosities in the X-ray -- radio luminosity plane in Figure \ref{fig:lxlr}. For comparison, we also plot black hole (black points) and neutron star LMXBs (grey circles), as well as supergiant HMXB (grey squares), and other outbursting BeXRBs (blue octagons and purple squares), taken from \citet{vandeneijnden2021} and \citet{gallo18}. LS V +44 17 is shown in the red octagons, where we stress that the lowest X-ray luminosity point consists of the two overlapping points from Observations 3 and 4. As shown by the light curve in Figure \ref{fig:lcs}, the radio counterpart is only seen at the highest X-ray luminosities, and shows what appears to be a scattered relation between its X-ray and radio luminosity. The final non-detection, in Observation 10, is obtained at relatively high X-ray luminosity, but evidently suffers from a lower sensitivity than the 9 other observations (Section \ref{sec:radiodet}). 

It is particularly noteworthy that the BeXRB 1A 0535+262 (blue octagons) is detected in the radio band at similar X-ray luminosities, but at a substantially lower radio luminosity: for instance, around $L_X = 10^{37}$ erg/s, LS V +44 17 is detected at a factor $\sim$4.5--7 brighter (Observation 7 and 8, respectively) than 1A 0535+262. The 1A 0535+262 flux density was measured at $6$ GHz, but the difference in observing band is unlikely to account for all of this difference; that would require a spectral index in the range $\alpha \approx 3$---$4$. 

To assess the presence and radio-brightness of an X-ray -- radio correlation for LS V +44 17 more quantitatively, we first consider the correlation between the two luminosities during the four observations with detections in both bands. For this radio-detected segment of the outburst monitoring, we measure a Pearson correlation coefficient of $r=0.64$, which, due to the small number of points and scatter, unexpectedly, has a low significance ($p=0.35$). The apparent presence of the coupling between X-ray and radio luminosity is, however, driven in large part by the systematic radio non-detections at the lower X-ray luminosities; their exclusion in calculating a correlation coefficient also severaly limits the dynamic range in X-ray luminosity that is considered. Therefore, to more properly use all observations to quantify the X-ray -- radio correlation, including radio upper limits, we also fit the standard X-ray binary correlation shape to the X-ray and radio data:

\begin{equation}
    L_R = \xi L_{R,0} \left(\frac{L_X}{L_{X,0}}\right)^{\beta} \text{ .}
\end{equation}

\noindent We highlight that by assuming this fitting model, we implicitly assume a correlation to be present between the two luminosities. Here, we use the average X-ray luminosity of the four observations with radio detections\footnote{Taking the average X-ray luminosity as the anchor $L_{X,0}$ minimizes the degeneracy between the slope and offset parameter in the fit.}, $L_{X,0} = 8.28\times10^{36}$ erg/s, and $L_{R,0} = 3.72\times10^{28}$ erg/s \citep[the latter taken from][]{gallo18}. In comparison to the radio luminosity, the uncertainties on the X-ray luminosity measurements are negligible: SNR$_{\rm X-ray}$ / SNR$_{\rm radio} > 10^2$ (see Table \ref{tab:data}). Therefore, we treat the X-ray luminosity as an independent variable with no uncertainties. We perform a model fit assuming uniform priors on the unit-less normalisation $\xi$ between $10^{-2}$ and $1$ and slope $\beta$ between $0$ and $4$. Specifically, we determine the combination of $\xi$ and $\beta$ that maximizes the joint probability of detecting the observed radio luminosities during Observations 6--9 and not detecting radio emission at the $3$-$\sigma$ level during the remaining epochs. We assume Gaussian noise in both cases: e.g., we maximize the multiplied probabilities of a radio luminosity as observed (detections) or below three times the observation RMS (non-detections) assuming a Gaussian distribution with the model radio luminosity and observation RMS as mean and standard deviation, respectively. With this approach we measure a slope $\beta = 1.25^{+0.66}_{-0.28}$ and a normalisation parameter $\xi = (2.5\pm0.3)\times10^{-2}$, where we report the parameter modes and their $16^{\rm th}$/$84^{\rm th}$ percentile. The fitted slope is significantly non-zero, i.e. positive, which is consistent with the presence of a positive coupling between X-ray and radio luminosity -- albeit with significant scatter in the radio-detected points. In Figure \ref{fig:lxlr}, the shaded pink area shows the region bounded by these limits on the slope and normalisation, while the red line shows the best fit correlation. 

In this correlation fit, we have assumed an anchor X-ray luminosity $L_{X,0}$ equal to the average $L_X$ during radio-detected observations; due to the radio faintness of neutron star BeXRBs, however, this value is a factor $\sim 4$ larger than the value of $2\times10^{36}$ erg/s used in the literature to compare subclasses of X-ray binaries \citep{gallo18}. We can therefore rescale the normalisation parameter $\xi$ to that X-ray luminosity, finding $\xi = (4.3\pm0.5)\times10^{-3}$ \citep[i.e. $\log \xi = -2.37 \pm 0.05$, in the notation used in][]{gallo18}, which corresponds to a radio luminosity normalisation of $\xi L_{R,0} = (1.6\pm0.2)\times10^{26}$ erg/s. This normalisation significantly exceeds the value measured for 1A 0535+262 and Swift J0243.6+6124 combined, e.g. $\xi = (2.5\pm0.3)\times10^{-3}$ \citep[][or $\log \xi = -2.61 \pm 0.05$; shown as the light blue region in Figure \ref{fig:lxlr}]{vandeneijnden2022}. This difference, confirming our earlier and more approximate comparison around $L_X \sim 10^{37}$ erg/s, corresponds to a factor $1.7$ difference in radio luminosity normalisation. As we will discuss more in the next section, we stress that this comparison with Swift J0243.6+6124 does not include the data from its X-ray re-brightening \citep{vandeneijnden2019}.

\section{Discussion}

In this paper, we have presented a VLA radio monitoring campaign of LS V +44 17 during its 2022/2023 outburst. While LS V +44 17 is a BeXRB persistently accreting at low levels, the monitored activity interval was its X-ray brightest period recorded to date. Our four radio detections at 10 GHz and two accompanying detections at 6 GHz make LS V +44 17 the third radio-detected neutron star BeXRB. The radio monitoring -- a combination of upper limits at low X-ray luminosity and detections at higher luminosity -- is consistent with a coupling between the radio emission and X-ray activity in a scattered fashion that is also consistent with the type of correlation seen in other X-ray binary classes. Around an X-ray luminosity of $10^{37}$ erg/s, its radio luminosity significantly exceeds that of 1A 0535+262, providing the first evidence of scatter between sources in their giant BeXRB outburst behaviour in radio. We note that all radio luminosities observed for LS V +44 17 fall below the limit obtained for GRO J1008-57 during both its periastron and giant outbursts \citep{vandeneijnden2022}. In this Discussion, we will first consider briefly the origin of the detected radio emission. We will then turn to the X-ray -- radio coupling, and continue under the assumption that the two luminosities are correlated: for that scenario, we will discuss the presence X-ray -- radio correlation scatter, considering both general X-ray binary and specific BeXRB explanations.

\subsection{The origin of radio counterpart of LS V +44 17}
\label{sec:41}

Through several lines of arguments that are discussed extensively in \citet{vandeneijnden2021,vandeneijnden2022}, the radio emission of LS V +44~17 is unlikely to originate from the Be-star. The transient nature of the radio emission and its coupling with the X-ray luminosity are instead consistent with a non-thermal scenario. Of the possible non-thermal mechanisms, emission from a propeller-driven outflow \citep{illarionov1975,romanova2008,parfrey2017}, either directly or from its interactions with e.g. a weak stellar wind, can be ruled out; the propeller regime occurs at low accretion rate, when the magnetospheric radius moves beyond the co-rotation radius, so that propeller-driven outflows should not become brighter at higher mass accretion rates\footnote{Moreover, LS V +44 17 may not be expected to move into a propeller regime, as its accretion disc could recombine into a cold, neutral state before its propeller transition (see also Section \ref{sec:lrdens}; \citealt{tsygankov2017}).}. Non-thermal radio emission from a collimated jet-type outflow, instead of a wider wind-type outflow can instead explain the observed emission: such a radio jet may be driven by neutron star magnetic field lines opened up by the accretion flow and therefore be powered by the NS’s magnetic field and spin \citep{parfrey2016,das2022}. Jet outflows were invoked by \citet{vandeneijnden2018} and \citet{vandeneijnden2022} to similarly explain the low-frequency emission of the two other radio-detected neutron star BeXRB in their giant outbursts. The presence of an X-ray -- radio coupling, as well as the region of the $L_X$--$L_R$ diagram inhabited by all three sources during their giant outbursts, remains consistent with a single origin for their transient radio emission. 

While all three BeXRBs inhabit this same region during their giant outbursts, the X-ray re-brightenings of Swift J0243.6+6124 remain an exception to that behaviour. During one of those X-ray re-brightenings, \citet{vandeneijnden2019} reported a period of coupled radio activity at luminosities significantly exceeding the giant outburst correlation at similar X-ray luminosity. While LS V +44 17 appears radio brighter than the two other radio-monitored giant BeXRB outbursts, it remains over an order of magnitude fainter than Swift J0243.6+6124 during its re-brightening. The origin for this difference remains unclear. We can similarly make a link to periastron passage monitoring: while the earliest X-ray re-brightenings of Swift J0243.6+6124 after its giant discovery outburst appeared to repeat on a time scale similar to its orbital time scale, their unexpectedly bright radio counterpart does not behave as seen in periastron radio monitoring of other BeXRBs: it is brighter than the limits on periastron radio emission in GRO J1008-57 \citep{vandeneijnden2022}, and also greatly exceeds our new limits for the radio emission of LS V +44 17 during the 2023 outburst peak that coincided with periastron (see e.g. Figure \ref{fig:lcs}). 

As an alternative to jet emission, \citet{chatzis2022} explored the scenario where an accretion-disc-driven outflow shocks with the weak Be-star wind, somewhat akin to the gamma-ray binary model where the pulsar wind and stellar wind shock to produce bright non-thermal emission \citep[e.g.][]{dubus2013}. The correlation between X-ray and radio emission predicted in this scenario ($\beta = 12/7 \approx 1.71$) is significantly steeper than that observed in the giant outbursts of 1A 0535+262 and Swift J0243.6+6124; it is, however, consistent within the upper $1$-$\sigma$ limit for LS V +44 17. The mechanism to launch a disc wind in the $L_X$ regime of the LS V +44 17 radio detection remains unclear; propeller and super-Eddington radiation-driven outflows \citep{shakura1973,ohsuga2011} occur at lower and higher $L_X$, respectively. Observationally, disc wind signatures have tentatively been detected in super-Eddington states \citep{vandeneijnden2019, koliopanos2016}, but no such evidence exists from X-ray grating spectra at sub-Eddington luminosities \citep{lapalombara2016, grinberg2017}.

We stress, however, that an absence of X-ray evidence for disc winds does not imply that no winds are launched: the presence of the bright early-type donor with emission lines complicates IR/optical/UV wind detection methods that have successfully probed winds in LMXBs \citep[see e.g.][]{xrbwinds}. The presence of thermal and magnetic winds therefore remains poorly understood in HMXB accretion discs. The recent launch of IXPE may have opened up a new avenue to probe the presence of disc winds in the X-rays; \textit{IXPE} observations of LS V +44 17 indicate the presence of a polarized but unpulsed component, consistent with a highly equatorial disc wind \citep[amongst other possibilities;][]{doroshenko2023}. Whether such a wind is present --- and can play the role of the more spherical disc wind assumed by \citet{chatzis2022} --- remains inconclusive with current evidence (the BeXRBs observed at lower X-ray luminosities, X Per and GRO J1008-57, also do not show similar polarization evidence for a disc wind, although that may be expected in thermal wind launch mechanisms that depend on the X-ray luminosity; \citealt{tsygankov2023,mushtukov2023}). Regardless, the \textit{IXPE} observations of LS V +44 17 highlight the value of including X-ray polarimetry in multi-band monitoring of transient XRBs, in particular in conjunction with searches for outflows in X-rays and radio.

\subsection{The $L_X$--$L_R$ scatter of BeXRBs: what affects jet luminosity?}
\label{sec:lrdens}

\subsubsection{The BeXRB scatter compared to LMXBs}

Our radio monitoring of LS V +44 17 reveals significant scatter in the $L_X$--$L_R$ relation of giant outbursts of BeXRBs as a source class. As noted in Section \ref{sec:lxlrcor}, its extrapolated 6-GHz radio luminosity is significantly higher than that of 1A 0535+262 around $L_X \approx 10^{37}$ erg/s by a factor $\sim$4.5--7. Fitting all datapoints (i.e. including radio non-detections) with a standard X-ray -- radio correlation function, we find a normalisation $\log \xi = -2.37\pm0.05$ for LS V +44 17 (for $L_{X,0} = 2\times10^{36}$ erg/s), compared to $\log \xi = -2.61 \pm 0.05$ for 1A 0535+262 and Swift J0243.6+6124 (during its 2017/2018 \textit{giant} outburst only) combined. While those values are just inconsistent at $1$-$\sigma$, the normalisation for all three BeXRBs are consistent at that level when fitted as three individual sources (due to the significantly increased uncertainties for the latter two when treating them separately). Therefore, the results presented in this paper provide evidence for the presence of scatter between sources in the $L_X$--$L_R$ relation of giant BeXRB outbursts, with LS V +44 17 radio brighter than the other two systems. This result comes in addition to the known scatter around their best-fit correlation for each of the sources individually, both within and between outbursts; comparing the Swift J0243.6+6124 between its 2017/2018 and 2023 outbursts reveals that during the rise of the latter outburst, radio emission is seen a factor three above the limit on radio emission at similar X-ray luminosity in the 2017/2018 outburst rise \citep{vandeneijnden2023}. The scatter \textit{between} sources revealed by LS V +44 17 differs from that within individual sources, however, as it represent a systematic shift of its X-ray -- radio correlation instead of excess variance around a shared relation. 

Before discussing scatter in the radio luminosity, we briefly consider the X-ray luminosity. The high pulse fraction of BeXRBs, reaching tens of per cent but varying with both energy and accretion rate, may affect the relation between the mean, measured X-ray luminosity and the accretion rate. It may, for instance, lead to a slight under-estimation of the accretion rate and thereby liberated gravitational energy. Such effects are, however, small, as even for a $100$\% pulsed BeXRB, the mean X-ray luminosity would only underestimate the peak by a factor two -- a factor much smaller than the multiple orders of magnitude traversed during a full outburst. Furthermore, comparing pulse studies of the three radio-detected BeXRBs \citep{sagalnik2023b,beri2021,wang2022b}, we do not find evidence for systematic differences between the magnitude of their pulse fractions, nor their behaviour as a function of energy or accretion rate. Therefore, in the remainder of this Section, we will focus on causes for scatter in the radio luminosity, instead.

Scatter in X-ray -- radio correlations are commonly seen in both black hole and neutron star LMXBs \citep[e.g.][]{gallo2014,gallo18}, as well as around the fundamental plane of black hole activity that includes supermassive black holes \citep[e.g.][]{merloni2003}. For LMXBs, \citet{gallo18} measure a similar level of statistical scatter for both black hole and neutron star systems. The scatter in the black hole sample is significantly increased by the presence of a radio-loud and radio-quiet correlation, which are combined in the above analysis, for which links to system inclination \citep{motta2018} and radio spectral shape \citep{espinasse2018} have been suggested. Even within the radio-loud track, however, differences in radio luminosity up to a factor $\sim 5$ can be seen in Figure \ref{fig:lxlr}; the factors amongst neutron star LMXBs are even larger. The origin of these latter two forms of scatter remain poorly understood, although differences between e.g. specific sub-classes of neutron star LMXBs do not appear to correlate with their radio brightness \citep[e.g.][]{gallo18,Gusinskaia2020b,vandeneijnden2021}. \citet{Gusinskaia2020b} furthermore list and consider a range of other explanations, including orbital period \citep[see also][]{tetarenko2018}, neutron star spin \citep[see also][]{migliari2011,russell2018}, inclination, and disc winds \citep[see also][]{tudor17}: none of these possibilities were found to be capable of explaining the observed scatter. With three radio-detected BeXRBs, it is challenging to search for definitive dependence on properties such as inclination or period. However, whatever origin underlies the LMXB scatter may be similarly at work in BeXRBs. 

\subsubsection{BeXRB-specific explanations for the scatter}

Alternatively, we can turn to the distinct properties of BeXRBs compared to LMXBs in the search for explanations of their $L_X$--$L_R$ scatter. Firstly, BeXRB jets may be powered by different mechanisms than those in LMXBs, due to their order of magnitude stronger magnetic fields ($> 10^{12}$ G vs. $<10^9$ G, typically, with exceptions from a handful of strongly-magnetized neutron stars in LMXBs). In the jet models by \citet{parfrey2016} and \citet{das2022}, the jet is powered by a combination of the NSs spin $P$ and magnetic field strength $B$ that differs from models typically assumed for LMXBs \citep[e.g.][]{blandford1982}. Specifically, if we assume that the jet radio luminosity scales with jet power to the power 1.4 \citep[as typically assumed for black hole jets;][]{blandford1979,markoff2001,falcke1996}, those neutron star jet models predict that $L_R~\propto~P^{-14/5} B^{6/5}$. Evidently, the strongest dependence is on neutron star spin, with a faster spin yielding a more powerful and brighter jet. Comparing the three radio-detected neutron star BeXRBs, we see that Swift J0243.6+6124 is the fastest spinning ($\sim 9.8$ sec), while 1A 0535+262 and LS V +44 17 spin significantly slower ($\sim 103$ sec and $\sim 205$ sec). Specifically, the above scaling would imply a difference in jet luminosity of $\sim 5\times10^3$ between Swift J0243.6+6124 and LS V +44 17, which is in strong contrast with their similar regions in the X-ray -- radio luminosity diagram. In other words, our monitoring of LS V +44 17 further strengthens the conclusion from \citet{vandeneijnden2022} that proposed theoretical scalings with spin and magnetic fields do not quantitatively reproduce the observed BeXRB differences within this subclass --- despite correctly predicting the qualitative difference between the weakly-magnetized neutron star LMXBs and the strongly-magnetized HMXBs. 

Similarly, the data from the three BeXRBs combined argues against a simplistic scaling with magnetic field, where a weaker magnetic field and therefore smaller magnetospheric radius would lead to a larger gravitational power reservoir for the jet to tap into: recent studies of LS V +44 17 \citep{sagalnik2023b} argue against earlier claims of a cyclotron line around 30 keV and instead suggest a strong, ${\sim}10^{13}$ G magnetic field, which would put the source at the high end of known BeXRB field strengths \citep{staubert2019}. Even if the cyclotron line in LS V +44 17 is confirmed, the large difference in field strength between Swift J0243.6+6124 and 1A 0535+262 \citep{kong2022}, combined with their consistent $L_R$ normalizations, argues against this scenario. 

In constrast to LMXBs, the jets in BeXRBs may also encounter low-density ambient stellar wind material that could play a role in regulating the radio luminosity of BeXRB jets. LS V +44 17 and 1A 0535+262 are both examples of persistent BeXRBs, which accrete at low-levels ($10^{34} - 10^{35}$ erg/s) in between rare outbursts \citep{reig1999,ferrigno2013,rothschild2013}. Two possible explanations for this behaviour have been proposed in the literature; originally, it was proposed that this X-ray emission could originate from low-level accretion from a weak spherical Be-star wind \citep{reig1999,ferrigno2013,sguera2023}. More recently, the possibility of accretion from a cold, re-combined disc was proposed \citep{tsygankov2017, sagalnik2023b} for systems with a spin $>100$ seconds. Either mechanism may operate in LS V +44 17 and 1A 0535+262 \citep[see also e.g.][]{doroshenko2023}, but the literature does not currently contain direct observables to distinguish between the scenarios.

In the case of low-level accretion from a stellar wind, such spherically distributed material would presumably remain in place when a giant outburst is triggered and a jet is launched. This jet can crash into and sweep up such material, and/or may be re-collimated by the wind, all leading to (re-collimation) shocks and therefore more effective conversion of jet kinetic power into radiative power. Alternatively, the wind may also bend the jet direction, as considered in calculations for the interactions between the more powerful stellar winds and jets in microquasars \citep{yoon2015,boschramon2016,barkov2022,lopezmiralles2022}, that are radio bright even among the stellar-mass black hole population: examples are radio-bright sources with an extreme intra-binary or surrounding density such as Cyg X-3 and SS 433. 

The density of such ambient wind material can differ strongly between sources, depending on the properties of their stellar wind and orbit. For a source with a quiescent X-ray luminosity $L_{X,q}$, the assumption of Bondi-Hoyle-Lyttleton accretion from a stellar wind implies \citep[following][]{frank2002} a stellar mass loss rate of 

\begin{equation}
\dot{M}_{\rm wind} = \frac{a^2 v_{\infty}^4 L_{X,q}}{\eta c^2 G^2 M^2_{\rm NS}} \text{ ,}
\label{eq:mdot}
\end{equation}

\noindent where we have assumed that $L_{X,q} = \eta\dot{M}_{\rm acc} c^2$, that the wind has reached its terminal velocity $v_\infty$ at the orbital distance, and $a$ is the orbital separation. Inserting this mass loss rate into the radial density profile of the wind implies, assuming mass conservation ($\rho_{\rm wind}(r) = \dot{M}_{\rm wind}/4\pi r^2 v_{\rm wind}(r)$) and jet-wind interactions at a height above the plane much smaller than the orbital size, that the jet encounters a density

\begin{equation}
\rho_{\rm wind}(r=a) = \frac{v_{\infty}^3 L_{X,q}}{4\pi\eta c^2 G^2 M^2_{\rm NS}} \text{ .}
\label{eq:rho}
\end{equation}

\noindent Together, these two relations imply that two sources with a very similar orbital separation and quiescent X-ray luminosity --- such as LS V +44 15 and 1A 0535+262 --- may have significantly different ambient densities. The driving factor is the wind’s velocity: if $v_\infty$ is twice as high in LS V +44 15, while its mass loss rate is sixteen times higher, both sources would have similar persistent X-ray luminosities while the jet in LS V +44 15 would encounter an eight times higher ambient density. 

More quantitatively, for a source with an orbital separation of $10^{13}$ cm, $L_{X, q} = 10^{34}$ erg/s, $\eta=0.1$, and a wind velocity of $500$ km/s, the Be star would need to lose mass at a rate of $3\times10^{-8}$ $M_{\odot}$/yr; the jet would then encounter an ambient number density of approximately $2\times10^{7}$ cm$^{-3}$, assuming a pure hydrogen wind. At twice the wind velocity, those values would increase to $4.8\times10^{-7}$ $M_{\odot}$/yr and $1.6\times10^{8}$ cm$^{-3}$. Such values of the mass loss rate would be high for a Be star \citep{krticka2014}, in particular the latter case, possibly pushing beyond the expected range for this stellar type \citep{vink2000}. 

The interaction between the jet and an ambient wind at these densities, apt for BeXRBs, is poorly explored. Jet-stellar wind interactions have been observed and modelled in the black hole system Cyg X-1 specifically\citep[e.g.][where the wind density from the O9.7Iab donor encountered by the jet is of the order $5\times10^9$ cm$^{-3}$ for their assumed parameters]{zdziarski2012} and microquasars generally \citep{yoon2015,boschramon2016,barkov2022,lopezmiralles2022}. No such studies, on the other hand, have been performed for the significantly weaker jets of neutron star BeXRBs interacting with the wind densities of Be stars, rather than the supergiant OB stars that typically reside in smaller orbits in SgXBs. Therefore, it is currently unclear what quantitative effects jet-wind interactions in BeXRBs could have on the jet's geometry and luminosity. However, the significantly brighter radio emission of the subset of radio-detected neutron star SgXBs \citep{vandeneijnden2021}, hosting OB supergiant donors with stronger stellar winds, would fit into the broader picture where the stellar wind (density) plays a role in regulating the observed jet brightness. As stated before, jet re-collimation shocks and bending, as for instance calculated and parameterized for microquasars \citep[e.g.][]{boschramon2016}, are likely much more effective for the weaker neutron star HMXB jets. In fact, particularly when interacting with the strong stellar winds in SgXBs, one may imagine cases where the weak jet only barely escapes or does not extend beyond the wind regions that are optically thick to radio emission, possibly accounting for radio non-detections of a subset of neutron star SgXBs \citep{vandeneijnden2021}. A detailed treatment of this question, including the effects of absorption, is beyond the scope of this paper, but warrants further investigation in future work.

\section{Summary and conclusions}

In this paper, we have presented radio and X-ray monitoring of the 2022/2023 outburst of the persistent neutron star BeXRB LS V +44 17. Its brightest ever period of accretion activity displayed two X-ray peaks, while radio emission was only detected during the second peak. The lack of detectable radio emission during the first, fainter outburst peak can be explained by an X-ray -- radio correlation in the source. During the second outburst peak, radio emission is detected during four observations, broadly tracking the X-rays and peaking around a comparable date. Comparing the X-ray and radio evolution in the X-ray -- radio luminosity diagram, we observe a scattered correlation. These results make LS V +44 17 only the third neutron star BeXRB with a radio counterpart, that is similarly consistent with a jet origin. Its location in the X-ray -- radio luminosity diagram is similar to the other two neutron star BeXRBs, Swift J0243.6+6124 and 1A 0535+262, with its scattered correlation showing a consistent slope. However, LS V +44 17 appears to be radio brighter at comparable X-ray luminosity, providing the first evidence for scatter between sources in the BeXRB X-ray -- radio plane. Such scatter is similarly seen in LMXBs, where it remains poorly understood; while the same, unknown origin may underlie the scatter seen in BeXRBs, we explicitly discuss a number of BeXRB-specific explanations. The nature of the scatter is in contrast with the predictions for scalings between jet power, spin, and magnetic field, made by neutron star jets models operating in the strong-magnetic-field regime \citep{parfrey2016,das2022}. Alternatively, the ambient stellar wind density may play a role in determining the exact radio jet luminosity. This latter scenario requires more detailed consideration in future work to assess, as well as a larger number of radio-detected BeXRBs to unravel the driving factors of their radio luminosity. 

\section{Acknowledgements}
We thank the referee for a constructive review. For the purpose of open access, the author has applied a Creative Commons Attribution (CC-BY) licence to any Author Accepted Manuscript version arising from this submission. JvdE acknowledges a Warwick Astrophysics prize post-doctoral fellowship made possible thanks to a generous philanthropic donation, and was supported by a Lee Hysan Junior Research Fellowship awarded by St. Hilda’s College, Oxford, during part of this work. ARE is supported by the European Space Agency (ESA) Research Fellowship. TDR acknowledges support as an INAF IAF fellow. GRS is supported by NSERC Discovery Grant RGPIN-2021-0400. GRS respectfully acknowledges that they perform the majority of their research from Treaty 6 territory, a traditional gathering place for diverse Indigenous peoples, including the Cree, Blackfoot, M\'etis, Nakota Sioux, Iroquois, Dene, Ojibway/Saulteaux/Anishinaabe, Inuit, and many others whose histories, languages, and cultures continue to influence our vibrant community. This research has made use of the MAXI data provided by RIKEN, JAXA, and the MAXI team. This research has made use of NASA's Astrophysics Data System Bibliographic Services. 

\section*{Data Availability}

A GitHub reproduction repository will be made public upon acceptance of this paper at \url{https://github.com/jvandeneijnden/LSV44\_17\_RepPackage} and will contain CASA analysis scripts, X-ray spectra and model files, as well as analysis and plotting scripts. An archived and stable release of this reproduction repository at the time of paper accceptance is also available on Zenodo via DOI \href{https://zenodo.org/doi/10.5281/zenodo.10063866}{10.5281/zenodo.10063866}. Due to size constraints, the unreduced VLA observations can be accessed publicly via the NRAO data archive (\url{data.nrao.edu}) starting 12 months after the observing date (i.e. in January and February 2024) by searching for project code {22B-051}. Earlier access to the unreduced VLA data can be granted upon reasonable request to the main author, especially to be used in conjunction with the GitHub/Zenodo reproduction repository. 

\input{main.bbl}


\appendix

\section{X-ray spectral parameters}
\label{app}

\begin{table*}
\caption{The spectral parameters for all ten fitted \textit{NICER} spectra. The model fitted to the data was \textsc{tbabs*(bbody + cutoffpl + gauss)}. Fitting all spectra jointly, the absorption column was found to be $N_H = (5.00\pm0.18)\times10^{21}$ cm$^{-2}$. The asterisk (*) indicates that a parameter was frozen to its maximum value as the fit was insensitive to the parameter.}
\label{tab:app}
{\renewcommand{\arraystretch}{1.5}\begin{tabular}{llllll}
\hline \hline
Parameter & 5203610110 & 5203610113 & 5203610115 & 5203610117 & 5203610121  \\
\hline
$kT_{\rm BB}$ [keV] & $0.50^{+0.03}_{-0.03}$ & $0.59^{+0.03}_{-0.03}$ & $0.67^{+0.03}_{-0.03}$ & $0.68^{+0.03}_{-0.03}$ & $0.49^{+0.02}_{-0.02}$ \\

$N_{\rm BB}$& $\left(9.19^{+2.20}_{-1.90}\right)\times10^{-4}$ & $\left(9.41^{+3.49}_{-2.90}\right)\times10^{-4}$ & $\left(1.15^{+0.42}_{-0.32}\right)\times10^{-3}$ & $\Gamma$ $\left(1.31^{+0.46}_{-0.35}\right)\times10^{-3}$ & $\left(2.51^{+0.40}_{-0.35}\right)\times10^{-3}$ \\

$\Gamma$ & $-0.54^{+0.06}_{-0.07}$ & $-0.65^{+0.11}_{-0.14}$ & $-0.83^{+0.16}_{-0.21}$ & $-0.94^{+0.17}_{-0.23}$ & $-0.75^{+0.10}_{-0.11}$ \\

$E_{\rm cut}$ [keV] & $4.24^{+0.19}_{-0.19}$ & $3.76^{+0.26}_{-0.26}$ & $3.24^{+0.27}_{-0.28}$ & $3.14^{+0.27}_{-0.28}$ & $3.88^{+0.25}_{-0.25}$ \\

$N_{\rm cpl}$ & $\left(6.98^{+0.39}_{-0.43}\right)\times10^{-2}$ & $\left(4.96^{+0.53}_{-0.58}\right)\times10^{-2}$ & $\left(3.16^{+0.51}_{-0.58}\right)\times10^{-2}$ & $\left(2.79^{+0.50}_{-0.56}\right)\times10^{-2}$ & $\left(7.84^{+0.65}_{-0.70}\right)\times10^{-2}$ \\

$E_{\rm Gauss}$ [keV] & $6.42^{+0.01}_{-0.01}$ & $6.41^{+0.01}_{-0.01}$ & $6.45^{+0.02}_{-0.02}$ & $6.40^{+0.05}$ & $6.43^{+0.01}_{-0.02}$ \\
$\sigma_{\rm Gauss}$ [keV]& $0.06^{+0.02}_{-0.02}$ & $0.05^{+0.03}_{-0.03}$ & $0.02^{+0.04}_{-0.02}$ & $0.10^{+0.09}_{-0.10}$ & $0.10^{+0.02}_{-0.02}$ \\

$N_{\rm Gauss}$ & $\left(1.37^{+0.16}_{-0.14}\right)\times10^{-3}$ & $\left(9.12^{+1.61}_{-1.48}\right)\times10^{-4}$ & $\left(4.38^{+1.14}_{-1.00}\right)\times10^{-4}$ & $\left(4.33^{+1.97}_{-1.93}\right)\times10^{-4}$ & $\left(3.39^{+0.42}_{-0.40}\right)\times10^{-3}$ \\ \hline \hline

Parameter & 5203610124 & 5203610134 & 5203610139 & 5203610148 & 5203610155 \\ \hline

$kT_{\rm BB}$ [keV] & $0.49^{+0.02}_{-0.02}$ & $0.13^{+0.03}_{-0.03}$ & $0.14^{+0.04}_{-0.03}$ & $0.36^{+0.02}_{-0.02}$ & $0.40^{+0.02}_{-0.02}$ \\
$N_{\rm BB}$ & $\left(4.26^{+0.46}_{-0.42}\right)\times10^{-3}$ & $\left(8.31^{+25.0}_{-4.26}\right)\times10^{-3}$ & $\left(7.84^{+17.5}_{-3.86}\right)\times10^{-3}$ & $\left(4.99^{+0.55}_{-0.53}\right)\times10^{-3}$ & $\left(5.40^{+0.76}_{-0.71}\right)\times10^{-3}$ \\
$\Gamma$ & $-0.65^{+0.08}_{-0.09}$ & $0.76^{+0.01}_{-0.01}$ & $0.75^{+0.01}_{-0.03}$ & $0.24^{+0.06}_{-0.07}$ & $-0.17^{+0.12}_{-0.13}$ \\
$E_{\rm cut}$ [keV] & $4.55^{+0.28}_{-0.27}$ & $500.00$* & $500.00$* & $12.78^{+1.85}_{-1.53}$ & $7.40^{+1.30}_{-1.04}$ \\
$N_{\rm cpl}$& $\left(1.09^{+0.08}_{-0.08}\right)\times10^{-1}$ & $\left(6.31^{0.03}_{-0.05}\right)\times10^{-1}$ & $\left(5.96^{+0.07}_{-0.11}\right)\times10^{-1}$ & $\left(3.34^{+0.16}_{-0.17}\right)\times10^{-1}$ & $\left(1.90^{+0.17}_{-0.18}\right)\times10^{-1}$ \\
$E_{\rm Gauss}$ [keV] & $6.43^{+0.01}_{-0.01}$ & $6.48^{+0.02}_{-0.02}$ & $6.47^{+0.02}_{-0.02}$ & $6.51^{+0.02}_{-0.02}$ & $6.47^{+0.03}_{-0.03}$ \\
$\sigma_{\rm Gauss}$ [keV] & $0.11^{+0.02}_{-0.02}$ & $0.48^{+0.05}_{-0.04}$ & $0.50^{+0.05}_{-0.04}$ & $0.32^{+0.03}_{-0.03}$ & $0.22^{+0.06}_{-0.05}$ \\
$N_{\rm Gauss}$ & $\left(5.33^{+0.52}_{-0.51}\right)\times10^{-3}$ & $\left(3.06^{+0.25}_{-0.23}\right)\times10^{-2}$ & $\left(2.98^{+0.26}_{-0.20}\right)\times10^{-2}$ & $\left(1.67^{+0.15}_{-0.14}\right)\times10^{-2}$ & $\left(1.07^{+0.21}_{-0.18}\right)\times10^{-2}$ \\ 
\hline
\end{tabular}}\\
\end{table*}

\bsp	
\label{lastpage}

\begin{thebibliography}{}
\makeatletter
\relax
\def\mn@urlcharsother{\let\do\@makeother \do\$\do\&\do\#\do\^\do\_\do\%\do\~}
\def\mn@doi{\begingroup\mn@urlcharsother \@ifnextchar [ {\mn@doi@} {\mn@doi@[]}}
\def\mn@doi@[#1]#2{\def\@tempa{#1}\ifx\@tempa\@empty \href {http://dx.doi.org/#2} {doi:#2}\else \href {http://dx.doi.org/#2} {#1}\fi \endgroup}
\def\mn@eprint#1#2{\mn@eprint@#1:#2::\@nil}
\def\mn@eprint@arXiv#1{\href {http://arxiv.org/abs/#1} {{\tt arXiv:#1}}}
\def\mn@eprint@dblp#1{\href {http://dblp.uni-trier.de/rec/bibtex/#1.xml} {dblp:#1}}
\def\mn@eprint@#1:#2:#3:#4\@nil{\def\@tempa {#1}\def\@tempb {#2}\def\@tempc {#3}\ifx \@tempc \@empty \let \@tempc \@tempb \let \@tempb \@tempa \fi \ifx \@tempb \@empty \def\@tempb {arXiv}\fi \@ifundefined {mn@eprint@\@tempb}{\@tempb:\@tempc}{\expandafter \expandafter \csname mn@eprint@\@tempb\endcsname \expandafter{\@tempc}}}

\bibitem[\protect\citeauthoryear{{Arnaud}}{{Arnaud}}{1996}]{xspecref}
{Arnaud} K.~A.,  1996, in {Jacoby} G.~H.,  {Barnes} J.,  eds,  Astronomical Society of the Pacific Conference Series Vol. 101, Astronomical Data Analysis Software and Systems V. p.~17

\bibitem[\protect\citeauthoryear{{Bailer-Jones}, {Rybizki}, {Fouesneau}, {Demleitner}  \& {Andrae}}{{Bailer-Jones} et~al.}{2021}]{Bailer-Jones2021}
{Bailer-Jones} C.~A.~L.,  {Rybizki} J.,  {Fouesneau} M.,  {Demleitner} M.,   {Andrae} R.,  2021, \mn@doi [\aj] {10.3847/1538-3881/abd806}, \href {https://ui.adsabs.harvard.edu/abs/2021AJ....161..147B} {161, 147}

\bibitem[\protect\citeauthoryear{{Barkov} \& {Bosch-Ramon}}{{Barkov} \& {Bosch-Ramon}}{2022}]{barkov2022}
{Barkov} M.~V.,  {Bosch-Ramon} V.,  2022, \mn@doi [\mnras] {10.1093/mnras/stab3609}, \href {https://ui.adsabs.harvard.edu/abs/2022MNRAS.510.3479B} {510, 3479}

\bibitem[\protect\citeauthoryear{{Barthelmy} et~al.,}{{Barthelmy} et~al.}{2005}]{batreference}
{Barthelmy} S.~D.,  et~al., 2005, \mn@doi [\ssr] {10.1007/s11214-005-5096-3}, \href {https://ui.adsabs.harvard.edu/abs/2005SSRv..120..143B} {120, 143}

\bibitem[\protect\citeauthoryear{{Beri} et~al.,}{{Beri} et~al.}{2021}]{beri2021}
{Beri} A.,  et~al., 2021, \mn@doi [\mnras] {10.1093/mnras/staa3254}, \href {https://ui.adsabs.harvard.edu/abs/2021MNRAS.500..565B} {500, 565}

\bibitem[\protect\citeauthoryear{{Blandford} \& {K{\"o}nigl}}{{Blandford} \& {K{\"o}nigl}}{1979}]{blandford1979}
{Blandford} R.~D.,  {K{\"o}nigl} A.,  1979, \mn@doi [\apj] {10.1086/157262}, \href {https://ui.adsabs.harvard.edu/abs/1979ApJ...232...34B} {232, 34}

\bibitem[\protect\citeauthoryear{{Blandford} \& {Payne}}{{Blandford} \& {Payne}}{1982}]{blandford1982}
{Blandford} R.~D.,  {Payne} D.~G.,  1982, \mn@doi [\mnras] {10.1093/mnras/199.4.883}, \href {https://ui.adsabs.harvard.edu/abs/1982MNRAS.199..883B} {199, 883}

\bibitem[\protect\citeauthoryear{{Bosch-Ramon} \& {Barkov}}{{Bosch-Ramon} \& {Barkov}}{2016}]{boschramon2016}
{Bosch-Ramon} V.,  {Barkov} M.~V.,  2016, \mn@doi [\aap] {10.1051/0004-6361/201628564}, \href {https://ui.adsabs.harvard.edu/abs/2016A&A...590A.119B} {590, A119}

\bibitem[\protect\citeauthoryear{{CASA Team} et~al.,}{{CASA Team} et~al.}{2022}]{casa2022}
{CASA Team} et~al., 2022, \mn@doi [\pasp] {10.1088/1538-3873/ac9642}, \href {https://ui.adsabs.harvard.edu/abs/2022PASP..134k4501C} {134, 114501}

\bibitem[\protect\citeauthoryear{{Caballero} et~al.,}{{Caballero} et~al.}{2013}]{caballero2013}
{Caballero} I.,  et~al., 2013, \mn@doi [\apjl] {10.1088/2041-8205/764/2/L23}, \href {https://ui.adsabs.harvard.edu/abs/2013ApJ...764L..23C} {764, L23}

\bibitem[\protect\citeauthoryear{{Casares}, {Negueruela}, {Rib{\'o}}, {Ribas}, {Paredes}, {Herrero}  \& {Sim{\'o}n-D{\'\i}az}}{{Casares} et~al.}{2014}]{casares2014}
{Casares} J.,  {Negueruela} I.,  {Rib{\'o}} M.,  {Ribas} I.,  {Paredes} J.~M.,  {Herrero} A.,   {Sim{\'o}n-D{\'\i}az} S.,  2014, \mn@doi [\nat] {10.1038/nature12916}, \href {https://ui.adsabs.harvard.edu/abs/2014Natur.505..378C} {505, 378}

\bibitem[\protect\citeauthoryear{{Chatzis}, {Petropoulou}  \& {Vasilopoulos}}{{Chatzis} et~al.}{2022}]{chatzis2022}
{Chatzis} M.,  {Petropoulou} M.,   {Vasilopoulos} G.,  2022, \mn@doi [\mnras] {10.1093/mnras/stab3098}, \href {https://ui.adsabs.harvard.edu/abs/2022MNRAS.509.2532C} {509, 2532}

\bibitem[\protect\citeauthoryear{{Coley} et~al.,}{{Coley} et~al.}{2023}]{coley2023}
{Coley} J.~B.,  et~al., 2023, The Astronomer's Telegram, \href {https://ui.adsabs.harvard.edu/abs/2023ATel15907....1C} {15907, 1}

\bibitem[\protect\citeauthoryear{{Corbel}, {Nowak}, {Fender}, {Tzioumis}  \& {Markoff}}{{Corbel} et~al.}{2003}]{corbel03}
{Corbel} S.,  {Nowak} M.~A.,  {Fender} R.~P.,  {Tzioumis} A.~K.,   {Markoff} S.,  2003, \mn@doi [\aap] {10.1051/0004-6361:20030090}, \href {http://cdsads.u-strasbg.fr/abs/2003A%26A...400.1007C} {400, 1007}

\bibitem[\protect\citeauthoryear{{Coriat} et~al.,}{{Coriat} et~al.}{2011}]{coriat2011}
{Coriat} M.,  et~al., 2011, \mn@doi [\mnras] {10.1111/j.1365-2966.2011.18433.x}, \href {https://ui.adsabs.harvard.edu/abs/2011MNRAS.414..677C} {414, 677}

\bibitem[\protect\citeauthoryear{{Das}, {Porth}  \& {Watts}}{{Das} et~al.}{2022}]{das2022}
{Das} P.,  {Porth} O.,   {Watts} A.~L.,  2022, \mn@doi [\mnras] {10.1093/mnras/stac1817}, \href {https://ui.adsabs.harvard.edu/abs/2022MNRAS.515.3144D} {515, 3144}

\bibitem[\protect\citeauthoryear{{Doroshenko} et~al.,}{{Doroshenko} et~al.}{2023}]{doroshenko2023}
{Doroshenko} V.,  et~al., 2023, \mn@doi [arXiv e-prints] {10.48550/arXiv.2306.02116}, \href {https://ui.adsabs.harvard.edu/abs/2023arXiv230602116D} {p. arXiv:2306.02116}

\bibitem[\protect\citeauthoryear{{Dubus}}{{Dubus}}{2013}]{dubus2013}
{Dubus} G.,  2013, \mn@doi [\aapr] {10.1007/s00159-013-0064-5}, \href {https://ui.adsabs.harvard.edu/abs/2013A&ARv..21...64D} {21, 64}

\bibitem[\protect\citeauthoryear{{Espinasse} \& {Fender}}{{Espinasse} \& {Fender}}{2018}]{espinasse2018}
{Espinasse} M.,  {Fender} R.,  2018, \mn@doi [\mnras] {10.1093/mnras/stx2467}, \href {https://ui.adsabs.harvard.edu/abs/2018MNRAS.473.4122E} {473, 4122}

\bibitem[\protect\citeauthoryear{{Falcke} \& {Biermann}}{{Falcke} \& {Biermann}}{1996}]{falcke1996}
{Falcke} H.,  {Biermann} P.~L.,  1996, \mn@doi [\aap] {10.48550/arXiv.astro-ph/9506138}, \href {https://ui.adsabs.harvard.edu/abs/1996A&A...308..321F} {308, 321}

\bibitem[\protect\citeauthoryear{{Ferrigno}, {Farinelli}, {Bozzo}, {Pottschmidt}, {Klochkov}  \& {Kretschmar}}{{Ferrigno} et~al.}{2013}]{ferrigno2013}
{Ferrigno} C.,  {Farinelli} R.,  {Bozzo} E.,  {Pottschmidt} K.,  {Klochkov} D.,   {Kretschmar} P.,  2013, \mn@doi [\aap] {10.1051/0004-6361/201321053}, \href {https://ui.adsabs.harvard.edu/abs/2013A&A...553A.103F} {553, A103}

\bibitem[\protect\citeauthoryear{{Finger} \& {Camero-Arranz}}{{Finger} \& {Camero-Arranz}}{2010}]{finger2010}
{Finger} M.~H.,  {Camero-Arranz} A.,  2010, The Astronomer's Telegram, \href {https://ui.adsabs.harvard.edu/abs/2010ATel.2537....1F} {2537, 1}

\bibitem[\protect\citeauthoryear{{Fortin}, {Garc{\'\i}a}, {Simaz Bunzel}  \& {Chaty}}{{Fortin} et~al.}{2023}]{fortin2023}
{Fortin} F.,  {Garc{\'\i}a} F.,  {Simaz Bunzel} A.,   {Chaty} S.,  2023, \mn@doi [\aap] {10.1051/0004-6361/202245236}, \href {https://ui.adsabs.harvard.edu/abs/2023A&A...671A.149F} {671, A149}

\bibitem[\protect\citeauthoryear{{Franchini} \& {Martin}}{{Franchini} \& {Martin}}{2019}]{franchini2019}
{Franchini} A.,  {Martin} R.~G.,  2019, \mn@doi [\apjl] {10.3847/2041-8213/ab3920}, \href {https://ui.adsabs.harvard.edu/abs/2019ApJ...881L..32F} {881, L32}

\bibitem[\protect\citeauthoryear{{Frank}, {King}  \& {Raine}}{{Frank} et~al.}{2002}]{frank2002}
{Frank} J.,  {King} A.,   {Raine} D.~J.,  2002, {Accretion Power in Astrophysics: Third Edition}

\bibitem[\protect\citeauthoryear{{Gaishin}, {Brumback}, {Kesler}, {Morrow}  \& {Wanink}}{{Gaishin} et~al.}{2023}]{gaishin2023}
{Gaishin} A.,  {Brumback} M.,  {Kesler} E.,  {Morrow} N.,   {Wanink} M.,  2023, The Astronomer's Telegram, \href {https://ui.adsabs.harvard.edu/abs/2023ATel16014....1G} {16014, 1}

\bibitem[\protect\citeauthoryear{{Gallo}, {Fender}  \& {Pooley}}{{Gallo} et~al.}{2003}]{gallo03}
{Gallo} E.,  {Fender} R.~P.,   {Pooley} G.~G.,  2003, \mn@doi [\mnras] {10.1046/j.1365-8711.2003.06791.x}, \href {http://adsabs.harvard.edu/abs/2003MNRAS.344...60G} {344, 60}

\bibitem[\protect\citeauthoryear{{Gallo} et~al.,}{{Gallo} et~al.}{2014}]{gallo2014}
{Gallo} E.,  et~al., 2014, \mn@doi [\mnras] {10.1093/mnras/stu1599}, \href {https://ui.adsabs.harvard.edu/abs/2014MNRAS.445..290G} {445, 290}

\bibitem[\protect\citeauthoryear{{Gallo}, {Degenaar}  \& {van den Eijnden}}{{Gallo} et~al.}{2018}]{gallo18}
{Gallo} E.,  {Degenaar} N.,   {van den Eijnden} J.,  2018, \mn@doi [\mnras] {10.1093/mnrasl/sly083}, \href {http://cdsads.u-strasbg.fr/abs/2018MNRAS.478L.132G} {478, L132}

\bibitem[\protect\citeauthoryear{{Gehrels} et~al.,}{{Gehrels} et~al.}{2004}]{gehrels2004}
{Gehrels} N.,  et~al., 2004, \mn@doi [\apj] {10.1086/422091}, \href {https://ui.adsabs.harvard.edu/abs/2004ApJ...611.1005G} {611, 1005}

\bibitem[\protect\citeauthoryear{{Gendreau} et~al.,}{{Gendreau} et~al.}{2016}]{gendreau2016}
{Gendreau} K.~C.,  et~al., 2016, in {den Herder} J.-W.~A.,  {Takahashi} T.,   {Bautz} M.,  eds,  Society of Photo-Optical Instrumentation Engineers (SPIE) Conference Series Vol. 9905, Space Telescopes and Instrumentation 2016: Ultraviolet to Gamma Ray. p. 99051H, \mn@doi{10.1117/12.2231304}

\bibitem[\protect\citeauthoryear{{Grinberg} et~al.,}{{Grinberg} et~al.}{2017}]{grinberg2017}
{Grinberg} V.,  et~al., 2017, \mn@doi [\aap] {10.1051/0004-6361/201731843}, \href {https://ui.adsabs.harvard.edu/abs/2017A&A...608A.143G} {608, A143}

\bibitem[\protect\citeauthoryear{{Gusinskaia} et~al.,}{{Gusinskaia} et~al.}{2020}]{Gusinskaia2020b}
{Gusinskaia} N.~V.,  et~al., 2020, \mn@doi [\mnras] {10.1093/mnras/stz3460}, \href {https://ui.adsabs.harvard.edu/abs/2020MNRAS.492.1091G} {492, 1091}

\bibitem[\protect\citeauthoryear{{HI4PI Collaboration} et~al.,}{{HI4PI Collaboration} et~al.}{2016}]{2016A&A...594A.116H}
{HI4PI Collaboration} et~al., 2016, \mn@doi [\aap] {10.1051/0004-6361/201629178}, \href {https://ui.adsabs.harvard.edu/abs/2016A&A...594A.116H} {594, A116}

\bibitem[\protect\citeauthoryear{{Hannikainen}, {Hunstead}, {Campbell-Wilson}  \& {Sood}}{{Hannikainen} et~al.}{1998}]{hannikainen98}
{Hannikainen} D.~C.,  {Hunstead} R.~W.,  {Campbell-Wilson} D.,   {Sood} R.~K.,  1998, \aap, \href {https://ui.adsabs.harvard.edu/abs/1998A&A...337..460H} {337, 460}

\bibitem[\protect\citeauthoryear{{Harrison} et~al.,}{{Harrison} et~al.}{2013}]{harrison2012}
{Harrison} F.~A.,  et~al., 2013, \mn@doi [\apj] {10.1088/0004-637X/770/2/103}, \href {https://ui.adsabs.harvard.edu/abs/2013ApJ...770..103H} {770, 103}

\bibitem[\protect\citeauthoryear{{Holder}}{{Holder}}{2023}]{veritas2023}
{Holder} J.,  2023, \mn@doi [arXiv e-prints] {10.48550/arXiv.2308.12214}, \href {https://ui.adsabs.harvard.edu/abs/2023arXiv230812214H} {p. arXiv:2308.12214}

\bibitem[\protect\citeauthoryear{{Illarionov} \& {Sunyaev}}{{Illarionov} \& {Sunyaev}}{1975}]{illarionov1975}
{Illarionov} A.~F.,  {Sunyaev} R.~A.,  1975, \aap, \href {https://ui.adsabs.harvard.edu/abs/1975A&A....39..185I} {39, 185}

\bibitem[\protect\citeauthoryear{{Janssens}, {Shenar}, {Degenaar}, {Bodensteiner}, {Sana}, {Audenaert}  \& {Frost}}{{Janssens} et~al.}{2023}]{janssens2023}
{Janssens} S.,  {Shenar} T.,  {Degenaar} N.,  {Bodensteiner} J.,  {Sana} H.,  {Audenaert} J.,   {Frost} A.~J.,  2023, \mn@doi [arXiv e-prints] {10.48550/arXiv.2308.08642}, \href {https://ui.adsabs.harvard.edu/abs/2023arXiv230808642J} {p. arXiv:2308.08642}

\bibitem[\protect\citeauthoryear{{Koliopanos} \& {Vasilopoulos}}{{Koliopanos} \& {Vasilopoulos}}{2018}]{koliopanos2016}
{Koliopanos} F.,  {Vasilopoulos} G.,  2018, \mn@doi [\aap] {10.1051/0004-6361/201731623}, \href {https://ui.adsabs.harvard.edu/abs/2018A&A...614A..23K} {614, A23}

\bibitem[\protect\citeauthoryear{{Kong} et~al.,}{{Kong} et~al.}{2022}]{kong2022}
{Kong} L.-D.,  et~al., 2022, \mn@doi [\apjl] {10.3847/2041-8213/ac7711}, \href {https://ui.adsabs.harvard.edu/abs/2022ApJ...933L...3K} {933, L3}

\bibitem[\protect\citeauthoryear{{Krimm} et~al.,}{{Krimm} et~al.}{2013}]{krimm2013}
{Krimm} H.~A.,  et~al., 2013, \mn@doi [\apjs] {10.1088/0067-0049/209/1/14}, \href {https://ui.adsabs.harvard.edu/abs/2013ApJS..209...14K} {209, 14}

\bibitem[\protect\citeauthoryear{{Krti{\v{c}}ka}}{{Krti{\v{c}}ka}}{2014}]{krticka2014}
{Krti{\v{c}}ka} J.,  2014, \mn@doi [\aap] {10.1051/0004-6361/201321980}, \href {https://ui.adsabs.harvard.edu/abs/2014A&A...564A..70K} {564, A70}

\bibitem[\protect\citeauthoryear{{K{\"u}hnel} et~al.,}{{K{\"u}hnel} et~al.}{2017}]{kuhnel2017}
{K{\"u}hnel} M.,  et~al., 2017, \mn@doi [\aap] {10.1051/0004-6361/201629131}, \href {https://ui.adsabs.harvard.edu/abs/2017A&A...607A..88K} {607, A88}

\bibitem[\protect\citeauthoryear{{Kumari}, {Pal}, {Mandal}  \& {Manna}}{{Kumari} et~al.}{2023}]{kumari2023}
{Kumari} S.,  {Pal} S.,  {Mandal} M.,   {Manna} A.,  2023, The Astronomer's Telegram, \href {https://ui.adsabs.harvard.edu/abs/2023ATel15913....1K} {15913, 1}

\bibitem[\protect\citeauthoryear{{La Palombara}, {Sidoli}, {Esposito}, {Tiengo}  \& {Mereghetti}}{{La Palombara} et~al.}{2012}]{lapalombara2012}
{La Palombara} N.,  {Sidoli} L.,  {Esposito} P.,  {Tiengo} A.,   {Mereghetti} S.,  2012, \mn@doi [\aap] {10.1051/0004-6361/201118221}, \href {https://ui.adsabs.harvard.edu/abs/2012A&A...539A..82L} {539, A82}

\bibitem[\protect\citeauthoryear{{La Palombara}, {Sidoli}, {Pintore}, {Esposito}, {Mereghetti}  \& {Tiengo}}{{La Palombara} et~al.}{2016}]{lapalombara2016}
{La Palombara} N.,  {Sidoli} L.,  {Pintore} F.,  {Esposito} P.,  {Mereghetti} S.,   {Tiengo} A.,  2016, \mn@doi [\mnras] {10.1093/mnrasl/slw020}, \href {https://ui.adsabs.harvard.edu/abs/2016MNRAS.458L..74L} {458, L74}

\bibitem[\protect\citeauthoryear{{Laplace}, {Mihara}, {Moritani}, {Nakajima}, {Takagi}, {Makishima}  \& {Santangelo}}{{Laplace} et~al.}{2017}]{laplace17}
{Laplace} E.,  {Mihara} T.,  {Moritani} Y.,  {Nakajima} M.,  {Takagi} T.,  {Makishima} K.,   {Santangelo} A.,  2017, \mn@doi [\aap] {10.1051/0004-6361/201629373}, \href {http://cdsads.u-strasbg.fr/abs/2017A%26A...597A.124L} {597, A124}

\bibitem[\protect\citeauthoryear{{L{\'o}pez-Miralles}, {Perucho}, {Mart{\'\i}}, {Migliari}  \& {Bosch-Ramon}}{{L{\'o}pez-Miralles} et~al.}{2022}]{lopezmiralles2022}
{L{\'o}pez-Miralles} J.,  {Perucho} M.,  {Mart{\'\i}} J.~M.,  {Migliari} S.,   {Bosch-Ramon} V.,  2022, \mn@doi [\aap] {10.1051/0004-6361/202142968}, \href {https://ui.adsabs.harvard.edu/abs/2022A&A...661A.117L} {661, A117}

\bibitem[\protect\citeauthoryear{{Mandal} et~al.,}{{Mandal} et~al.}{2023}]{mandal2023}
{Mandal} M.,  et~al., 2023, The Astronomer's Telegram, \href {https://ui.adsabs.harvard.edu/abs/2023ATel15848....1M} {15848, 1}

\bibitem[\protect\citeauthoryear{{Markoff}, {Falcke}  \& {Fender}}{{Markoff} et~al.}{2001}]{markoff2001}
{Markoff} S.,  {Falcke} H.,   {Fender} R.,  2001, \mn@doi [\aap] {10.1051/0004-6361:20010420}, \href {https://ui.adsabs.harvard.edu/abs/2001A&A...372L..25M} {372, L25}

\bibitem[\protect\citeauthoryear{{Martin} \& {Franchini}}{{Martin} \& {Franchini}}{2019}]{martin2019}
{Martin} R.~G.,  {Franchini} A.,  2019, \mn@doi [\mnras] {10.1093/mnras/stz2250}, \href {https://ui.adsabs.harvard.edu/abs/2019MNRAS.489.1797M} {489, 1797}

\bibitem[\protect\citeauthoryear{{Martin}, {Nixon}, {Armitage}, {Lubow}  \& {Price}}{{Martin} et~al.}{2014}]{martin14}
{Martin} R.~G.,  {Nixon} C.,  {Armitage} P.~J.,  {Lubow} S.~H.,   {Price} D.~J.,  2014, \mn@doi [\apjl] {10.1088/2041-8205/790/2/L34}, \href {http://cdsads.u-strasbg.fr/abs/2014ApJ...790L..34M} {790, L34}

\bibitem[\protect\citeauthoryear{{Massi} \& {Kaufman Bernad{\'o}}}{{Massi} \& {Kaufman Bernad{\'o}}}{2008}]{massi2008}
{Massi} M.,  {Kaufman Bernad{\'o}} M.,  2008, \mn@doi [\aap] {10.1051/0004-6361:20077567}, \href {https://ui.adsabs.harvard.edu/abs/2008A&A...477....1M} {477, 1}

\bibitem[\protect\citeauthoryear{{Matsuoka} et~al.,}{{Matsuoka} et~al.}{2009}]{maxiref}
{Matsuoka} M.,  et~al., 2009, \mn@doi [\pasj] {10.1093/pasj/61.5.999}, \href {https://ui.adsabs.harvard.edu/abs/2009PASJ...61..999M} {61, 999}

\bibitem[\protect\citeauthoryear{{Merloni}, {Heinz}  \& {di Matteo}}{{Merloni} et~al.}{2003}]{merloni2003}
{Merloni} A.,  {Heinz} S.,   {di Matteo} T.,  2003, \mn@doi [\mnras] {10.1046/j.1365-2966.2003.07017.x}, \href {https://ui.adsabs.harvard.edu/abs/2003MNRAS.345.1057M} {345, 1057}

\bibitem[\protect\citeauthoryear{{Migliari} \& {Fender}}{{Migliari} \& {Fender}}{2006}]{migliari06}
{Migliari} S.,  {Fender} R.~P.,  2006, \mn@doi [\mnras] {10.1111/j.1365-2966.2005.09777.x}, \href {http://adsabs.harvard.edu/abs/2006MNRAS.366...79M} {366, 79}

\bibitem[\protect\citeauthoryear{{Migliari}, {Miller-Jones}  \& {Russell}}{{Migliari} et~al.}{2011}]{migliari2011}
{Migliari} S.,  {Miller-Jones} J.~C.~A.,   {Russell} D.~M.,  2011, \mn@doi [\mnras] {10.1111/j.1365-2966.2011.18868.x}, \href {https://ui.adsabs.harvard.edu/abs/2011MNRAS.415.2407M} {415, 2407}

\bibitem[\protect\citeauthoryear{{Monageng}, {McBride}, {Coe}, {Steele}  \& {Reig}}{{Monageng} et~al.}{2017}]{monageng17}
{Monageng} I.~M.,  {McBride} V.~A.,  {Coe} M.~J.,  {Steele} I.~A.,   {Reig} P.,  2017, \mn@doi [\mnras] {10.1093/mnras/stw2354}, \href {http://cdsads.u-strasbg.fr/abs/2017MNRAS.464..572M} {464, 572}

\bibitem[\protect\citeauthoryear{{Morii} et~al.,}{{Morii} et~al.}{2010}]{morii2010}
{Morii} M.,  et~al., 2010, The Astronomer's Telegram, \href {https://ui.adsabs.harvard.edu/abs/2010ATel.2527....1M} {2527, 1}

\bibitem[\protect\citeauthoryear{{Moritani} et~al.,}{{Moritani} et~al.}{2013}]{moritani13}
{Moritani} Y.,  et~al., 2013, \mn@doi [\pasj] {10.1093/pasj/65.4.83}, \href {http://cdsads.u-strasbg.fr/abs/2013PASJ...65...83M} {65, 83}

\bibitem[\protect\citeauthoryear{{Motta}, {Casella}  \& {Fender}}{{Motta} et~al.}{2018}]{motta2018}
{Motta} S.~E.,  {Casella} P.,   {Fender} R.~P.,  2018, \mn@doi [\mnras] {10.1093/mnras/sty1440}, \href {https://ui.adsabs.harvard.edu/abs/2018MNRAS.478.5159M} {478, 5159}

\bibitem[\protect\citeauthoryear{{Mushtukov} et~al.,}{{Mushtukov} et~al.}{2023}]{mushtukov2023}
{Mushtukov} A.~A.,  et~al., 2023, \mn@doi [\mnras] {10.1093/mnras/stad1961}, \href {https://ui.adsabs.harvard.edu/abs/2023MNRAS.524.2004M} {524, 2004}

\bibitem[\protect\citeauthoryear{{Nakajima} et~al.,}{{Nakajima} et~al.}{2022}]{nakajima2022}
{Nakajima} M.,  et~al., 2022, The Astronomer's Telegram, \href {https://ui.adsabs.harvard.edu/abs/2022ATel15835....1N} {15835, 1}

\bibitem[\protect\citeauthoryear{{Ohsuga} \& {Mineshige}}{{Ohsuga} \& {Mineshige}}{2011}]{ohsuga2011}
{Ohsuga} K.,  {Mineshige} S.,  2011, \mn@doi [\apj] {10.1088/0004-637X/736/1/2}, \href {https://ui.adsabs.harvard.edu/abs/2011ApJ...736....2O} {736, 2}

\bibitem[\protect\citeauthoryear{{Okazaki} \& {Negueruela}}{{Okazaki} \& {Negueruela}}{2001}]{okazaki01}
{Okazaki} A.~T.,  {Negueruela} I.,  2001, \mn@doi [\aap] {10.1051/0004-6361:20011083}, \href {http://cdsads.u-strasbg.fr/abs/2001A%26A...377..161O} {377, 161}

\bibitem[\protect\citeauthoryear{{Okazaki}, {Hayasaki}  \& {Moritani}}{{Okazaki} et~al.}{2013}]{okazaki2013}
{Okazaki} A.~T.,  {Hayasaki} K.,   {Moritani} Y.,  2013, \mn@doi [\pasj] {10.1093/pasj/65.2.41}, \href {https://ui.adsabs.harvard.edu/abs/2013PASJ...65...41O} {65, 41}

\bibitem[\protect\citeauthoryear{{Pal} et~al.,}{{Pal} et~al.}{2023}]{pal2023}
{Pal} S.,  et~al., 2023, The Astronomer's Telegram, \href {https://ui.adsabs.harvard.edu/abs/2023ATel15868....1P} {15868, 1}

\bibitem[\protect\citeauthoryear{{Palmer} \& {Swift/BAT Team}}{{Palmer} \& {Swift/BAT Team}}{2023}]{palmer2023}
{Palmer} D.~M.,  {Swift/BAT Team} 2023, The Astronomer's Telegram, \href {https://ui.adsabs.harvard.edu/abs/2023ATel15880....1P} {15880, 1}

\bibitem[\protect\citeauthoryear{{Parfrey} \& {Tchekhovskoy}}{{Parfrey} \& {Tchekhovskoy}}{2017}]{parfrey2017}
{Parfrey} K.,  {Tchekhovskoy} A.,  2017, \mn@doi [\apjl] {10.3847/2041-8213/aa9c85}, \href {https://ui.adsabs.harvard.edu/abs/2017ApJ...851L..34P} {851, L34}

\bibitem[\protect\citeauthoryear{{Parfrey}, {Spitkovsky}  \& {Beloborodov}}{{Parfrey} et~al.}{2016}]{parfrey2016}
{Parfrey} K.,  {Spitkovsky} A.,   {Beloborodov} A.~M.,  2016, \mn@doi [\apj] {10.3847/0004-637X/822/1/33}, \href {http://adsabs.harvard.edu/abs/2016ApJ...822...33P} {822, 33}

\bibitem[\protect\citeauthoryear{{Pike} et~al.,}{{Pike} et~al.}{2023}]{pike2023}
{Pike} S.~N.,  et~al., 2023, \mn@doi [arXiv e-prints] {10.48550/arXiv.2306.16489}, \href {https://ui.adsabs.harvard.edu/abs/2023arXiv230616489P} {p. arXiv:2306.16489}

\bibitem[\protect\citeauthoryear{{Postnov}, {Mironov}, {Lutovinov}, {Shakura}, {Kochetkova}  \& {Tsygankov}}{{Postnov} et~al.}{2015}]{postnov2015}
{Postnov} K.~A.,  {Mironov} A.~I.,  {Lutovinov} A.~A.,  {Shakura} N.~I.,  {Kochetkova} A.~Y.,   {Tsygankov} S.~S.,  2015, \mn@doi [\mnras] {10.1093/mnras/stu2155}, \href {https://ui.adsabs.harvard.edu/abs/2015MNRAS.446.1013P} {446, 1013}

\bibitem[\protect\citeauthoryear{{Reig}}{{Reig}}{2011}]{reig2011}
{Reig} P.,  2011, \mn@doi [\apss] {10.1007/s10509-010-0575-8}, \href {https://ui.adsabs.harvard.edu/abs/2011Ap&SS.332....1R} {332, 1}

\bibitem[\protect\citeauthoryear{{Reig} \& {Roche}}{{Reig} \& {Roche}}{1999}]{reig1999}
{Reig} P.,  {Roche} P.,  1999, \mn@doi [\mnras] {10.1046/j.1365-8711.1999.02473.x}, \href {https://ui.adsabs.harvard.edu/abs/1999MNRAS.306..100R} {306, 100}

\bibitem[\protect\citeauthoryear{{Reig}, {Negueruela}, {Fabregat}, {Chato}  \& {Coe}}{{Reig} et~al.}{2005}]{reig2005}
{Reig} P.,  {Negueruela} I.,  {Fabregat} J.,  {Chato} R.,   {Coe} M.~J.,  2005, \mn@doi [\aap] {10.1051/0004-6361:20053124}, \href {https://ui.adsabs.harvard.edu/abs/2005A&A...440.1079R} {440, 1079}

\bibitem[\protect\citeauthoryear{{Remillard} et~al.,}{{Remillard} et~al.}{2022}]{remillard2019}
{Remillard} R.~A.,  et~al., 2022, \mn@doi [\aj] {10.3847/1538-3881/ac4ae6}, \href {https://ui.adsabs.harvard.edu/abs/2022AJ....163..130R} {163, 130}

\bibitem[\protect\citeauthoryear{{Rib{\'o}} et~al.,}{{Rib{\'o}} et~al.}{2017}]{ribo2017}
{Rib{\'o}} M.,  et~al., 2017, \mn@doi [\apjl] {10.3847/2041-8213/835/2/L33}, \href {https://ui.adsabs.harvard.edu/abs/2017ApJ...835L..33R} {835, L33}

\bibitem[\protect\citeauthoryear{{Rivinius}, {Klement}, {Chojnowski}, {Baade}, {Shepard}  \& {Hadrava}}{{Rivinius} et~al.}{2022}]{rivinius2022}
{Rivinius} T.,  {Klement} R.,  {Chojnowski} S.~D.,  {Baade} D.,  {Shepard} K.,   {Hadrava} P.,  2022, \mn@doi [arXiv e-prints] {10.48550/arXiv.2208.12315}, \href {https://ui.adsabs.harvard.edu/abs/2022arXiv220812315R} {p. arXiv:2208.12315}

\bibitem[\protect\citeauthoryear{{Romanova}, {Kulkarni}  \& {Lovelace}}{{Romanova} et~al.}{2008}]{romanova2008}
{Romanova} M.~M.,  {Kulkarni} A.~K.,   {Lovelace} R. V.~E.,  2008, \mn@doi [\apjl] {10.1086/527298}, \href {https://ui.adsabs.harvard.edu/abs/2008ApJ...673L.171R} {673, L171}

\bibitem[\protect\citeauthoryear{{Rothschild} et~al.,}{{Rothschild} et~al.}{2013}]{rothschild2013}
{Rothschild} R.,  et~al., 2013, \mn@doi [\apj] {10.1088/0004-637X/770/1/19}, \href {https://ui.adsabs.harvard.edu/abs/2013ApJ...770...19R} {770, 19}

\bibitem[\protect\citeauthoryear{{Rouco Escorial}, {Wijnands}, {van den Eijnden}, {Patruno}, {Degenaar}, {Parikh}  \& {Ootes}}{{Rouco Escorial} et~al.}{2020}]{rouco2020}
{Rouco Escorial} A.,  {Wijnands} R.,  {van den Eijnden} J.,  {Patruno} A.,  {Degenaar} N.,  {Parikh} A.,   {Ootes} L.~S.,  2020, \mn@doi [\aap] {10.1051/0004-6361/201936287}, \href {https://ui.adsabs.harvard.edu/abs/2020A&A...638A.152R} {638, A152}

\bibitem[\protect\citeauthoryear{{Russell}, {Degenaar}, {Wijnands}, {van den Eijnden}, {Gusinskaia}, {Hessels}  \& {Miller-Jones}}{{Russell} et~al.}{2018}]{russell2018}
{Russell} T.~D.,  {Degenaar} N.,  {Wijnands} R.,  {van den Eijnden} J.,  {Gusinskaia} N.~V.,  {Hessels} J.~W.~T.,   {Miller-Jones} J.~C.~A.,  2018, \mn@doi [\apjl] {10.3847/2041-8213/aaf4f9}, \href {https://ui.adsabs.harvard.edu/abs/2018ApJ...869L..16R} {869, L16}

\bibitem[\protect\citeauthoryear{{Salganik}, {Tsygankov}, {Doroshenko}, {Molkov}, {Lutovinov}, {Mushtukov}  \& {Poutanen}}{{Salganik} et~al.}{2023a}]{sagalnik2023b}
{Salganik} A.,  {Tsygankov} S.~S.,  {Doroshenko} V.,  {Molkov} S.~V.,  {Lutovinov} A.~A.,  {Mushtukov} A.~A.,   {Poutanen} J.,  2023a, \mn@doi [\mnras] {10.1093/mnras/stad2124}, \href {https://ui.adsabs.harvard.edu/abs/2023MNRAS.tmp.2056S} {}

\bibitem[\protect\citeauthoryear{{Salganik}, {Tsygankov}, {Lutovinov}  \& {Molkov}}{{Salganik} et~al.}{2023b}]{salganik2023}
{Salganik} A.,  {Tsygankov} S.~S.,  {Lutovinov} A.~A.,   {Molkov} S.~V.,  2023b, The Astronomer's Telegram, \href {https://ui.adsabs.harvard.edu/abs/2023ATel15874....1S} {15874, 1}

\bibitem[\protect\citeauthoryear{{S{\'a}nchez-Sierras} \& {Mu{\~n}oz-Darias}}{{S{\'a}nchez-Sierras} \& {Mu{\~n}oz-Darias}}{2020}]{xrbwinds}
{S{\'a}nchez-Sierras} J.,  {Mu{\~n}oz-Darias} T.,  2020, \mn@doi [\aap] {10.1051/0004-6361/202038406}, \href {https://ui.adsabs.harvard.edu/abs/2020A&A...640L...3S} {640, L3}

\bibitem[\protect\citeauthoryear{{Sguera}, {Sidoli}, {Bird}  \& {La Palombara}}{{Sguera} et~al.}{2023}]{sguera2023}
{Sguera} V.,  {Sidoli} L.,  {Bird} A.~J.,   {La Palombara} N.,  2023, \mn@doi [\mnras] {10.1093/mnras/stad1494}, \href {https://ui.adsabs.harvard.edu/abs/2023MNRAS.523.1192S} {523, 1192}

\bibitem[\protect\citeauthoryear{{Shakura} \& {Sunyaev}}{{Shakura} \& {Sunyaev}}{1973}]{shakura1973}
{Shakura} N.~I.,  {Sunyaev} R.~A.,  1973, \aap, \href {https://ui.adsabs.harvard.edu/abs/1973A&A....24..337S} {24, 337}

\bibitem[\protect\citeauthoryear{{Sidoli}}{{Sidoli}}{2017}]{Sidoli2017}
{Sidoli} L.,  2017, in XII Multifrequency Behaviour of High Energy Cosmic Sources Workshop (MULTIF2017). p.~52 (\mn@eprint {arXiv} {1710.03943}), \mn@doi{10.22323/1.306.0052}

\bibitem[\protect\citeauthoryear{{Staubert} et~al.,}{{Staubert} et~al.}{2019}]{staubert2019}
{Staubert} R.,  et~al., 2019, \mn@doi [\aap] {10.1051/0004-6361/201834479}, \href {https://ui.adsabs.harvard.edu/abs/2019A&A...622A..61S} {622, A61}

\bibitem[\protect\citeauthoryear{{Tetarenko} et~al.,}{{Tetarenko} et~al.}{2018}]{tetarenko2018}
{Tetarenko} A.~J.,  et~al., 2018, \mn@doi [\apj] {10.3847/1538-4357/aaa95a}, \href {https://ui.adsabs.harvard.edu/abs/2018ApJ...854..125T} {854, 125}

\bibitem[\protect\citeauthoryear{{Tsygankov}, {Lutovinov}  \& {Krivonos}}{{Tsygankov} et~al.}{2011}]{tsygankov2011}
{Tsygankov} S.,  {Lutovinov} A.,   {Krivonos} R.,  2011, The Astronomer's Telegram, \href {https://ui.adsabs.harvard.edu/abs/2011ATel.3137....1T} {3137, 1}

\bibitem[\protect\citeauthoryear{{Tsygankov}, {Krivonos}  \& {Lutovinov}}{{Tsygankov} et~al.}{2012}]{tsygankov2012}
{Tsygankov} S.~S.,  {Krivonos} R.~A.,   {Lutovinov} A.~A.,  2012, \mn@doi [\mnras] {10.1111/j.1365-2966.2012.20475.x}, \href {https://ui.adsabs.harvard.edu/abs/2012MNRAS.421.2407T} {421, 2407}

\bibitem[\protect\citeauthoryear{{Tsygankov}, {Wijnands}, {Lutovinov}, {Degenaar}  \& {Poutanen}}{{Tsygankov} et~al.}{2017a}]{tsygankov2017b}
{Tsygankov} S.~S.,  {Wijnands} R.,  {Lutovinov} A.~A.,  {Degenaar} N.,   {Poutanen} J.,  2017a, \mn@doi [\mnras] {10.1093/mnras/stx1255}, \href {https://ui.adsabs.harvard.edu/abs/2017MNRAS.470..126T} {470, 126}

\bibitem[\protect\citeauthoryear{{Tsygankov}, {Mushtukov}, {Suleimanov}, {Doroshenko}, {Abolmasov}, {Lutovinov}  \& {Poutanen}}{{Tsygankov} et~al.}{2017b}]{tsygankov2017}
{Tsygankov} S.~S.,  {Mushtukov} A.~A.,  {Suleimanov} V.~F.,  {Doroshenko} V.,  {Abolmasov} P.~K.,  {Lutovinov} A.~A.,   {Poutanen} J.,  2017b, \mn@doi [\aap] {10.1051/0004-6361/201630248}, \href {https://ui.adsabs.harvard.edu/abs/2017A&A...608A..17T} {608, A17}

\bibitem[\protect\citeauthoryear{{Tsygankov} et~al.,}{{Tsygankov} et~al.}{2023}]{tsygankov2023}
{Tsygankov} S.~S.,  et~al., 2023, \mn@doi [\aap] {10.1051/0004-6361/202346134}, \href {https://ui.adsabs.harvard.edu/abs/2023A&A...675A..48T} {675, A48}

\bibitem[\protect\citeauthoryear{{Tudor} et~al.,}{{Tudor} et~al.}{2017}]{tudor17}
{Tudor} V.,  et~al., 2017, \mn@doi [\mnras] {10.1093/mnras/stx1168}, \href {http://adsabs.harvard.edu/abs/2017MNRAS.470..324T} {470, 324}

\bibitem[\protect\citeauthoryear{{Verner}, {Ferland}, {Korista}  \& {Yakovlev}}{{Verner} et~al.}{1996}]{vern1996}
{Verner} D.~A.,  {Ferland} G.~J.,  {Korista} K.~T.,   {Yakovlev} D.~G.,  1996, \mn@doi [\apj] {10.1086/177435}, \href {https://ui.adsabs.harvard.edu/abs/1996ApJ...465..487V} {465, 487}

\bibitem[\protect\citeauthoryear{{Vink}, {de Koter}  \& {Lamers}}{{Vink} et~al.}{2000}]{vink2000}
{Vink} J.~S.,  {de Koter} A.,   {Lamers} H.~J.~G.~L.~M.,  2000, \mn@doi [\aap] {10.48550/arXiv.astro-ph/0008183}, \href {https://ui.adsabs.harvard.edu/abs/2000A&A...362..295V} {362, 295}

\bibitem[\protect\citeauthoryear{{Wang} et~al.,}{{Wang} et~al.}{2022}]{wang2022b}
{Wang} P.~J.,  et~al., 2022, \mn@doi [\apj] {10.3847/1538-4357/ac8230}, \href {https://ui.adsabs.harvard.edu/abs/2022ApJ...935..125W} {935, 125}

\bibitem[\protect\citeauthoryear{{Weisskopf} et~al.,}{{Weisskopf} et~al.}{2022}]{IXPEref}
{Weisskopf} M.~C.,  et~al., 2022, \mn@doi [Journal of Astronomical Telescopes, Instruments, and Systems] {10.1117/1.JATIS.8.2.026002}, \href {https://ui.adsabs.harvard.edu/abs/2022JATIS...8b6002W} {8, 026002}

\bibitem[\protect\citeauthoryear{{Wilms}, {Allen}  \& {McCray}}{{Wilms} et~al.}{2000}]{wilms2000}
{Wilms} J.,  {Allen} A.,   {McCray} R.,  2000, \mn@doi [\apj] {10.1086/317016}, \href {https://ui.adsabs.harvard.edu/abs/2000ApJ...542..914W} {542, 914}

\bibitem[\protect\citeauthoryear{{Yoon} \& {Heinz}}{{Yoon} \& {Heinz}}{2015}]{yoon2015}
{Yoon} D.,  {Heinz} S.,  2015, \mn@doi [\apj] {10.1088/0004-637X/801/1/55}, \href {https://ui.adsabs.harvard.edu/abs/2015ApJ...801...55Y} {801, 55}

\bibitem[\protect\citeauthoryear{{Zdziarski}}{{Zdziarski}}{2012}]{zdziarski2012}
{Zdziarski} A.~A.,  2012, \mn@doi [\mnras] {10.1111/j.1365-2966.2012.20754.x}, \href {https://ui.adsabs.harvard.edu/abs/2012MNRAS.422.1750Z} {422, 1750}

\bibitem[\protect\citeauthoryear{{van Kerkwijk} et~al.,}{{van Kerkwijk} et~al.}{1992}]{vankerkwijk1992}
{van Kerkwijk} M.~H.,  et~al., 1992, \mn@doi [\nat] {10.1038/355703a0}, \href {https://ui.adsabs.harvard.edu/abs/1992Natur.355..703V} {355, 703}

\bibitem[\protect\citeauthoryear{{van den Eijnden}, {Degenaar}, {Russell}, {Wijnands}, {Miller-Jones}, {Sivakoff}  \& {Hern{\'a}ndez Santisteban}}{{van den Eijnden} et~al.}{2018}]{vandeneijnden2018}
{van den Eijnden} J.,  {Degenaar} N.,  {Russell} T.~D.,  {Wijnands} R.,  {Miller-Jones} J.~C.~A.,  {Sivakoff} G.~R.,   {Hern{\'a}ndez Santisteban} J.~V.,  2018, \mn@doi [\nat] {10.1038/s41586-018-0524-1}, \href {https://ui.adsabs.harvard.edu/abs/2018Natur.562..233V} {562, 233}

\bibitem[\protect\citeauthoryear{{van den Eijnden} et~al.,}{{van den Eijnden} et~al.}{2019}]{vandeneijnden2019}
{van den Eijnden} J.,  et~al., 2019, \mn@doi [\mnras] {10.1093/mnras/stz1548}, \href {https://ui.adsabs.harvard.edu/abs/2019MNRAS.487.4355V} {487, 4355}

\bibitem[\protect\citeauthoryear{{van den Eijnden} et~al.,}{{van den Eijnden} et~al.}{2021}]{vandeneijnden2021}
{van den Eijnden} J.,  et~al., 2021, \mn@doi [\mnras] {10.1093/mnras/stab1995}, \href {https://ui.adsabs.harvard.edu/abs/2021MNRAS.507.3899V} {507, 3899}

\bibitem[\protect\citeauthoryear{{van den Eijnden}, {Degenaar}, {Russell}, {Miller-Jones}, {Rouco Escorial}, {Wijnands}, {Sivakoff}  \& {Hern{\'a}ndez Santisteban}}{{van den Eijnden} et~al.}{2022}]{vandeneijnden2022}
{van den Eijnden} J.,  {Degenaar} N.,  {Russell} T.~D.,  {Miller-Jones} J.~C.~A.,  {Rouco Escorial} A.,  {Wijnands} R.,  {Sivakoff} G.~R.,   {Hern{\'a}ndez Santisteban} J.~V.,  2022, \mn@doi [\mnras] {10.1093/mnras/stac2518}, \href {https://ui.adsabs.harvard.edu/abs/2022MNRAS.516.4844V} {516, 4844}

\bibitem[\protect\citeauthoryear{{van den Eijnden}, {Russell}, {Miller-Jones}, {Degenaar}, {Wijnands}, {Sivakoff}, {Rouco Escorial}  \& {Hernandez Santisteban}}{{van den Eijnden} et~al.}{2023}]{vandeneijnden2023}
{van den Eijnden} J.,  {Russell} T.,  {Miller-Jones} J.,  {Degenaar} N.,  {Wijnands} R.,  {Sivakoff} G.,  {Rouco Escorial} A.,   {Hernandez Santisteban} J.~V.,  2023, The Astronomer's Telegram, \href {https://ui.adsabs.harvard.edu/abs/2023ATel16101....1V} {16101, 1}

\makeatother
\end{thebibliography}
\end{document}